\documentstyle[12pt,epsfig]{article}  
\def \cos{\mathop{\rm cos}\nolimits}

\let \nm=\nonumber

\let \b=\bigskip
\let \d=\displaystyle
\let \q=\qquad

\newcommand{\beqn}{\begin{eqnarray*}}
\newcommand{\eeqn}{\end{eqnarray*}}
\newcommand{\beq}{\begin{eqnarray}}
\newcommand{\eeq}{\end{eqnarray}}
\newcommand{\eqn}{\begin{equation}}
\newcommand{\een}{\end{equation}}
\newcommand{\ba}{\begin{array}}
\newcommand{\ea}{\end{array}}

\begin{document}

\date{\ }
\title{$\rm QED_{2+1}$: Compton effect on Dirac-Landau electrons}
\author{ J. Mateos Guilarte \& M. de la Torre Mayado\\
 Departamento de F\'{\i}sica. Facultad de Ciencias\\  Universidad de Salamanca.
Salamanca 37008\\ SPAIN}
\maketitle

\begin{abstract}

Planar Quantum Electrodynamics is developed when  charged fermions are
under the influence of a constant and homogeneous
external magnetic field. We compute the cross-length for the scattering of
optical/ultraviolet
photons by Dirac-Landau electrons.

\end{abstract}

PACS: 12.20.-m, 11.10. Kk

\section{Introduction}

The second quantization method and the associated occupation number
formalism are basic pillars in quantum field theory.
Both in \lq\lq fundamental" elementary particle physics and condensed
matter systems one deals with many-particle
ensembles and the number of particles is not conserved \cite{Brown}. The
fermionic/bosonic Fock space  is built
out of the one particle Hilbert space of states by the antisymmetric/symmetric
tensor product: ${\cal F} =
\sum_{N=0}^{\infty} { \cal A} /{\cal S} \stackrel{N}{\otimes}L^2 ( {\bf
R}^n) $, where $N$ and $n$ are
respectively the number of particles and the dimension of the configuration
space. Usually, eigenfunctions of either the
momentum or position operators are taken as a \lq\lq basis" in $L^2({\bf
R}^n)$ and thus plane-waves or
$\delta$-functions are the one particle wave-functions on which the
procedure is based.

The interaction of photons, electrons and positrons when fermions are
subjected to a constant external
magnetic field is essentially described by quantum electrodynamics on a
plane orthogonal to the direction of the
magnetic field $\vec B$. Thus, here we shall discuss ${\rm QED}_{2+1}$
starting from a basis of Landau
states in $L^2({\bf R}^n)$, \cite{LandauNRQM}. Apart from providing an
example of the occupation number formalism 
not dependent on plane-wave states, the Dirac equation in an external
magnetic field also presents important novelties with
respect to the zero field case, e.g. spectral asymmetry.

 We shall
analyze the scattering of
photons by electrons under the action of an external homogeneous magnetic
field in perturbation theory. We set thus forth  a
physical situation closely related to that occurring in planar Hall devices
at very low temperatures and very high magnetic fields. To fit this in with the
relativistic
approach, we shall compare the theoretical outcome with electromagnetic
radiation scattering in samples where the
electron effective mass is very small. Such a Hall device could be the Hg
Cd Te MISFET
(metal-insulator-semiconductor-field-effect-transistor), see
\cite{R.Prange90}, which works at very low temperatures of around 1
degree Kelvin; this setting therefore also allows for a zero-temperature
field-theoretical treatment. Measures of the optic Hall angle in
this system, similar to those performed in high-Tc superconductors
\cite{Spiel}, might be addressed within the
 framework of planar QED.

The organization of the paper is as follows: in Section \S.2 we study the
Dirac equation in a homogeneous magnetic
field and quantize the Dirac-Landau field. Section \S.3 is devoted to
developing perturbation theory and its
application to the understanding of Compton scattering. Finally, in
Section \S.4 we compute the cross-length
for the scattering of optical/ultraviolet photons by Dirac-Landau electrons and
comment on several issues.

\section{Field expansion in Dirac-Landau states}

\subsection{ The Dirac equation in a homogeneous magnetic field}

The Dirac equation governing the quantum mechanics of a relativistic
charged particle of mass $m$ and spin $1/2$
is,
\eqn
\gamma^{\mu} \left( i \hbar \partial_{\mu} + {e\over c} A_{\mu}^{\rm ext} (x)
\right) \psi (x) =  m c \, \psi(x)  \label{2.1.1}
\een
if the fermion moves in a plane under a time-independet and homogeneous
magnetic field. Our conventions for the
metric, $2\times 2$ Dirac matrices and the like are explained  in appendix
A and planar Dirac fermions in external fields
are described in Reference \cite{Jackiw84}. We work
in the Weyl and Landau gauges where the three-vector external potential
reads as $A_{\mu}^{\rm ext} (x) = ( 0, \vec
A(x))$ and $A_1(x) = B x_2$, $A_2 (x) = 0$. This produces a constant and
uniform magnetic field $\vec B = - B
\vec k$ and the stationary states $ \psi_E (\vec x) e^{- {i E t \over
\hbar}}$ satisfy the spectral equation $H
\psi_E(\vec x) = E \psi_E(\vec x)$ for the Hamiltonian Dirac operator:
\beq
H &=& \left( \ba{cc} m c^2 & D \\ D^{\dag} & - m c^2 \ea \right)
\label{2.1.2} \\ D &=& - \sqrt{2 e B \hbar c} \
a^{\dag} \ , \ D^{\dag} = -\sqrt{2 e B \hbar c} \ a \nm
\eeq

The solution of the non-relativistic Landau problem in the plane is well
known, see \cite{LandauNRQM}. In terms of the annihilation and
creation operators $a$ and $a^{\dag}$,
\[
a = {1\over \sqrt{2} l} \left[ l^2 \left( { \partial \over \partial
x_2 }   - i {\partial \over \partial x_1 } \right) +  x_2  \right]  \ , \
a^{\dag } = {1 \over \sqrt{2} l} \left[ - l^2 \left(  {\partial \over
\partial x_2 } + i {\partial \over \partial x_1 } \right) +  x_2 \right]
 \]
that do not commute, $[a, a^{\dag}]= 1$, the Schr\"odinger operator is: $
H_S = \hbar \omega \left( a^{\dag } a +
{1\over 2} \right) $. $H_S$ commutes in the Landau gauge with $p_1 = - i
\hbar {\partial \over \partial x_1}$;
thus, there exists a complete set of eigenfunctions common to $H_S$ and
$p_1$ formed by products of Hermite
polynomials and plane-waves. $\omega = {e B \over m c}$ and $l^2 = {\hbar
\over m \omega }$ are the cyclotron
frequency and the magnetic length.

Given the $N = {eBA \over 2 \pi \hbar c}$ eigenfunctions of $H_S$,
\beq
& & \Phi_{n, p_1} (\vec x) = {1\over \sqrt{L_1} } e^{i p_1 x_1 \over \hbar}
\phi_{n, p_1} (x_2) \label{2.1.3} \\
& & \phi_{n, p_1} (x_2) = {1 \over \pi^{1/4} \sqrt{2^n n ! l}æ} H_n \left[
{x_2 + {p_1\overæ\hbar} l^2 \over l } \right] e^{-{1\over 2 l^2} (x_2 + {p_1 \over
\hbar } l^2 )^2 }\label{2.1.4}
\eeq
with the center of the orbit $ x^0_2 = -{ p_1 l^2 \over \hbar } $ located
in a rectangular enclosure of area $A=L_1 L_2$, one can easily find the
eigenfunctions for the Dirac operator
$H$. In (\ref{2.1.3}), (\ref{2.1.4}) $n = 0, 1, 2, \cdots$ label the Landau
levels, $p_1/\hbar = 2 \pi
q_1/\hbar$, so that $q_1 \in {\bf Z}$ is the \lq\lq discrete" momentum in
the $OX_1$-direction and $H_n[x]$ are
the Hermite polynomials. Therefore, the energy eigenvalues of $H$ are
\beq
E_n^{\pm} &=& \pm \sqrt{ 2 e B \hbar c \ n + m^2 c ^4} \ , \  n= 1,
2,\cdots \label{2.1.5} \\
E_0^{+} &=& + m c^2  \nm
\eeq
whereas the corresponding eigenspinors read
\beq
&& \psi_{n, p_1}^{\pm}(\vec x) = \sqrt{ E_n \pm m c^2 \over 2 E_n} \left( \ba{c}
\Phi_{n, p_1} (\vec x) \\ \\ {- E_n^0 \over {\pm E_n + m c^2} } \Phi_{n-1, p_1}
(\vec x) \ea \right) \, ,\, n=1, 2, \cdots \label{2.1.6}æ\\
&& \psi_{0, p_1} (\vec x) =   \left( \ba{c}
\Phi_{0, p_1} (\vec x) \\ \\ 0 \ea \right)  \nmæ
\eeq
if $E_n = |E_n^{\pm}|$ and $E_n^0 = + \sqrt{2 e B \hbar c \ n }$.
Accordingly, the Dirac-Landau spectral problem
shares the following properties with the non-relativistic counterpart: (1)
The spectrum is discrete and the
Dirac-Landau energy levels are labeled by a non-negative integer. (2) Each
energy level is degenerated
and the eigenvalues of $p_1$ characterize the degeneracy. Nevertheless,
there are two important differences: (1)
Infinite negative energy levels appear and we can talk of a
Dirac-Landau sea. (2) The spectrum shows a
spectral assymmetry associated with the fundamental or ground state; for
$n=0$ there are states with positive energy which are
not paired with others of negative energy. It is remarkable that the energy
of these states, which form what we shall
call the first Landau level, is independent of $B$. At the zero mass limit,
the first Landau level is spanned by
\lq\lq zero modes" of the Dirac operator.

For later convenience we introduce the notation:
\beq
&& \psi_{n, p_1}^{\pm}(\vec x) = {1\over \sqrt{L_1}} e^{i p_1 x_1 \over \hbar}
{\rm u}_{n p_1}^{\pm} (x_2) \ , \  {\rm u}_{n p_1}^{\pm} (x_2) = \sqrt{ E_n
\pm m c^2 \over 2
E_n} \left( \ba{c} \Phi_{n, p_1} (x_2) \\ \\ {- E_n^0 \over {\pm E_n + m c^2} }
\Phi_{n-1, p_1} (x_2) \ea \right) \, , \, n \neq 0 \nm \\
&&  \psi_{0, p_1} (\vec x) = {1\over
\sqrt{L_1}} e^{i p_1 x_1 \over \hbar} {\rm u}_{0 p_1}^+ (x_2) \ , \ {\rm
u}_{0 p_1}^+ (x_2) =   \left( \ba{c}
\Phi_{0, p_1} ( x_2) \\ \\ 0 \ea \right)  \label{2.1.7}
\eeq
and define the Fourier transform and its inverse for the two-spinors ${\rm
u}_{n p_1}^{\pm}$ and ${\rm u}_{o
p_1}^+$:
\beq
&& {\rm u}_{n p_1}^{\pm} (x_2) = {l \over 2 \pi  } \int_{-\infty}^{\infty}
{\cal U}_{n
p_1}^{\pm} (k_2) e^{i k_2 x_2  } d k_2 \ , \
{\rm u}_{0 p_1}^+  (x_2) = {l \over 2 \pi  } \int_{-\infty}^{\infty}  {\cal
U}_{0
p_1}^+ (k_2 ) e^{i k_2 x_2  } d k_2 \label{2.1.8} \\
&& {\cal U}_{n p_1}^{\pm} (k_2) = {1 \over l} \int_{-\infty}^{\infty} {\rm u}_{n
p_1}^{\pm} (x_2) e^{- i k_2 x_2 } d x_2 \ , \
{\cal U}_{0 p_1}^+  (k_2) = {1 \over l} \int_{-\infty}^{\infty}  {\rm u}_{0
p_1}^+  (x_2) e^{- i k_2 x_2 } d x_2 \nm
\eeq
Bearing in mind that $k_2 = p_2/\hbar$, we find
\eqn
 {\cal U}_{n p_1}^{\pm} (k_2)= \sqrt{ E_n \pm m c^2 \over 2 E_n} \left( \ba{c}
 {\varphi}_{n, p_1} (k_2) \\ \\ {- E_n^0 \over {\pm E_n + m c^2} }
 {\varphi}_{n-1, p_1} (k_2) \ea \right) \  ,\
 {\cal U}_{0 p_1}^+ (k_2) =  \left(
\ba{c}  {\varphi}_{0, p_1} ( k_2) \\ \\ 0 \ea \right) \label{2.1.9}
\een
where ${\varphi}_{n, p_1} (k_2)$ are the Fourier transforms of the
non-relativistic
Landau wave-functions:
\eqn
{\varphi}_{n, p_1} (k_2) = {1\over l} \int_{-\infty}^{\infty} \phi_{n, p_1}
(x_2) e^{-i k_2 x_2} d x_2
= {\sqrt{2 \pi} \over \pi^{1/4} i^n \sqrt{2^n n! l} }  H_n
\left[ {k_2 l }æ\right] e^{   i { p_1 k_2 l^2 \over \hbar}-{ k_2^2 l^2 \over
2 }} \label{2.1.10}
\een

\subsection{ The Dirac-Landau field}

Wave-particle duality, $E= \hbar \omega$, $\vec p = \hbar \vec k$ and
$\hbar e_{\rm Field} = e$, allows the
understanding of equation (\ref{2.1.1}) in classical field theory instead
of relativistic quantum mechanics. In
this framework the Dirac-Landau equation appears as the Euler-Lagrange
equation for the Lagrangian:
\eqn
{\cal L} = c \bar \psi (x) \left[ \gamma^{\mu} \left( i \hbar \partial_{\mu} +
{e\over c} A_{\mu}^{\rm ext} (x) \right) - m c \right] \psi (x) \label{2.2.11}
\een
Thus, in the Weyl gauge the field theoretical Dirac Hamiltonian is:
\eqn
H = \int d^2 x \left[  \psi^{\dag} (x) \left[ \vec \alpha (  - i \hbar c
\vec \nabla  + e
\vec A^{\rm ext} (x) ) + \beta \, m c^2 \right] \psi (x) \right] \label{2.2.12}
\een
where $\beta = \gamma^0$ and $ \alpha^j = \beta \gamma^j$, $j= 1, 2$, and
$\bar \psi (x) = \psi^{\dag} (x)
\gamma^0$.

In order to quantize this system, see \cite{Johnson}, we impose the
anti-commutation relations at equal times:
\beq
\{ \psi_{\alpha}(t,\vec x), \psi_{\beta}^{\dag}(t, \vec y) \} &=&
\delta_{\alpha \beta} \delta^{(2)} (\vec x -
\vec y) \label{2.2.13} \\
\{ \psi_{\alpha}(t,\vec x), \psi_{\beta} (t, \vec y) \} &=& \{
\psi_{\alpha}^{\dag}(t,\vec x),
\psi_{\beta}^{\dag}(t, \vec y) \} = 0  \nm
\eeq

The expansion of the Dirac field $\psi$ and its adjoint $\bar \psi$
\beq
\psi(x) &=& \psi^+ (x) + \psi^- (x) + \psi^0 (x) \nm \\ &=& \sum_{n, p_1} \left[
C_{n p_1} \psi_{n p_1}^{+} (\vec x) e^{- {i E_n t \over
\hbar}} + D_{n p_1}^{\dag} \psi_{n p_1}^{-} (\vec x) e^{i E_n t \over \hbar}
\right] + \sum_{p_1} A_{0 p_1} \psi_{0 p_1}(\vec x) e^{- {i E_0 t \over \hbar}}
\label{2.2.14}
\\
\bar{ \psi} (x) &=& \bar{\psi}^+ (x) + \bar{\psi}^- (x) + \bar{\psi}^0 (x)
\nm \\
&=& \sum_{n, p_1}
\left[ D_{n p_1} \bar{\psi}_{n p_1}^{-} (\vec x) e^{- {i E_n t \over
\hbar}} + C_{n p_1}^{\dag} \bar{\psi}_{n p_1}^{+} (\vec x) e^{i E_n t \over
\hbar}
\right] + \sum_{p_1} A_{0 p_1}^{\dag} \bar{\psi}_{0 p_1}(\vec x) e^{+ {i E_0 t
\over \hbar}}
\nm
\eeq
is compatible with (\ref{2.2.13}) if the coefficients $C_{n p_1}$, $D_{n
p_1}$ and $A_{0 p_1}$ are
operators satisfying the anticommutation relations:
\eqn
  \{ C_{n p_1}, C_{n',{ p'}_1}^{\dag} \} = \{ D_{n p_1}, D_{n',{
p'}_1}^{\dag} \}
 = \delta_{n, n'} \delta_{p_1, {p'}_1} \ , \
 \{ A_{0 p_1}, A_{0,{ p'}_1}^{\dag} \} = \delta_{p_1, {p'}_1} \label{2.2.15}
\een
and every other anticommutator between these operators vanishes.

The fermionic Fock-Landau space admits a basis  built out of the vacuum
\[
C_{n p_1} |0\rangle =  D_{n p_1} |0 \rangle =  A_{0 p_1} |0 \rangle = 0 \ ,
\ \forall n \, ,\, \forall p_1
\]
by the action of strings of creation operators
\beq
&& | \kappa_{n^{1} p_1^{1}}  \delta_{n^{1} p_1^{1}}  \alpha_{0 p_1^{1}}
\cdots \kappa_{n^{N} p_1^{N}}
\delta_{n^{N} p_1^{N}}  \alpha_{0 p_1^{N}}  \rangle
\nm \\
&& \propto  [C_{n^1 p_1^1}^{\dag}]^{\kappa_{n^{1} p_1^{1}} } [D_{n^1
p_1^1}^{\dag}]^{\delta_{n^{1} p_1^{1}} } [A_{0 p_1^1}^{\dag}]^{\alpha_{0
p_1^{1}} } \cdots [C_{n^N
p_1^N}^{\dag}]^{\kappa_{n^{N} p_1^{N}} }[D_{n^N
p_1^N}^{\dag}]^{\delta_{n^{N} p_1^{N}} }[A_{0
p_1^N}^{\dag}]^{\alpha_{0 p_1^{N}} } |0 \rangle
\label{2.2.16}
\eeq
where $\kappa_{n^{i} p_1^{i}}, \delta_{n^{i} p_1^{i}}, \alpha_{0 p_1^{i}}$
are $ 0$ or $1$, complying with
Fermi statistics. Therefore, the states of the basis are eigenvectors of
the number operators $N_{n p_1}^+ =
C_{n p_1}^{\dag} C_{n p_1}$, $N_{n p_1}^- = D_{n p_1}^{\dag} D_{n p_1}$ and
$N_{0 p_1} = A_{0 p_1}^{\dag} A_{0
p_1}$. From (\ref{2.2.15}) one can easily deduce that $C_{n p_1}$ and $C_{n
p_1}^{\dag}$ annihilate and create
respectively, electrons in the $n^{\underline {\rm th}}$ Landau level,
whereas $D_{n p_1}$ and $D_{n p_1}^{\dag}$
do the same job with positrons. $A_{0 p_1}$ and $A_{0 p_1}^{\dag}$ are
the annihilation and creation
operators of electrons occupying the first Landau level. The spectral
asymmetry of the Dirac-Landau operator
forbids the ocupation of the first Landau level by positrons in the second
quantization framework for this
system.

Our states of the basis are eigenvalues of the Hamiltonian, component one
of the momentum and charge operators,
which properly normal-ordered read:
\beq
H&=& \sum_{n=1}^{\infty} \sum_{p_1 = - \infty}^{\infty} E_n ( C_{n
p_1}^{\dag} C_{n p_1} + D_{n p_1}^{\dag} D_{n
p_1} ) + \sum_{p_1 = - \infty}^{\infty} E_0 \left(   A_{0 p_1}^{\dag} A_{0
p_1} - {1\over 2} \right) \nm \\
P_1&=& \sum_{n=1}^{\infty} \sum_{p_1 = - \infty}^{\infty} \hbar q_1 ( C_{n
p_1}^{\dag} C_{n p_1} + D_{n p_1}^{\dag}
D_{n p_1} ) + \sum_{p_1 = - \infty}^{\infty} \hbar q_1  A_{0 p_1}^{\dag}
A_{0 p_1}  \label{2.2.17} \\
Q&=& - e \sum_{n=1}^{\infty} \sum_{p_1 = - \infty}^{\infty}  ( C_{n
p_1}^{\dag} C_{n p_1} - D_{n p_1}^{\dag} D_{n
p_1} ) - e \sum_{p_1 = - \infty}^{\infty}  \left(   A_{0 p_1}^{\dag} A_{0
p_1} - {1\over 2} \right) \nm
\eeq
Notice that $\langle 0 | Q | 0 \rangle = {e\over 2} \eta (0)$, $\eta(s) =
\sum_{n \in {\bf Z} -\{0\}} {1\over n^s}
+ 1 $, because we do not have the $A_{0p_1}$ and
$A_{0p_1}^{\dag}$ normally ordered. This choice is made to
distinguish the Dirac-Landau sea from the normal situation where all the
particles have their anti-particles.
The states (\ref{2.2.16}) however, do not have definite spin because we are
working in the Landau gauge.

The anticommutation relations at different times are
\eqn
 \{ \psi(x), \psi(y) \} = \{ \bar{\psi}(x), \bar{\psi}(y) \} = 0 \ , \
 \{ \psi_{\alpha} (x), \bar{\psi}_{\beta}(y) \} = i S_{\alpha \beta} (x - y)
\label{2.2.18}
\een
where the $2\times 2$ matrix function $S (x) = S^+ (x) + S^- (x) + S^0 (x)$ is
\beq
& & i S^+ (x - y) = \sum_{n, p_1}  \psi_{n p_1}^+ (\vec x) \bar{\psi}_{n
p_1}^{+}
(\vec y) e^{- {i E_n (x_0 - y_0) \over \hbar c}} \nm \\
& & i S^- (x - y) = \sum_{n, p_1}  \psi_{n p_1}^- (\vec x) \bar{\psi}_{n
p_1}^{-}
(\vec y) e^{ {i E_n (x_0 - y_0) \over \hbar c}} \label{2.2.19} \\
& & i S^0 (x - y) = \sum_{ p_1}  \psi_{0 p_1}  (\vec x) \bar{\psi}_{0 p_1}
(\vec y) e^{- {i E_0 (x_0 - y_0) \over \hbar c}} \nm
\eeq

The fermion propagator in a magnetic field is the expectation value of the
time ordered product $T\{ \psi (x),
\bar{\psi} (y) \}$ at the vacuum state:
\beq
i S_F (x-y) &=& \langle 0| T\{ \psi (x), \bar{\psi} (y) \} | 0 \rangle
\label{2.2.20} \\
&=& i \left( \theta (x_0 - y_0) S^+ (x-y) - \theta(y_0-x_0) S^- (x-y)
\right) + i \theta(x_0-y_0) S^0 (x-y) \nm
\eeq
($\theta (x) $ is the step function, $\theta (x) = 1$ if $x>0$,
$\theta(x)=0$ if $x<0$).
Taking into account finite temperature effects requires that one must define the
propagator as:
\beq
i S_F(\beta;x-y) ={{\rm Tr} (e^{-\beta H} T\{\psi (x), \bar{\psi} (y) \}) \over
{\rm Tr} e^{-\beta H}}
\eeq
where $\beta$ is the inverse temperature. Temperature Green functions like
this can be
computed in the canonical formalism, see \cite{Rivers87}: $T\neq 0$ effects
are included by considering
a complex Minkowski time and (anti)-periodic fields in the imaginary
component of period $i\beta$. One considers a contour C between 0 and
$i\beta$ containing the real axis and defines the path ordering along C.
This amounts to doubling the fields: $\psi_1= \psi (x_0,\vec x) , \psi_2=
-\psi (x_0-i{\beta\over 2}, \vec x)$. If $\beta \rightarrow \infty $
$\psi_2$ decouples and , equivalently, only the vaccum state contributes
to $ S_F (x-y) $, which is given by (\ref{2.2.20}) . Thus, at very low
temperatures we reach a very good
approximation by considering $\beta=\infty$ QED.
\section{Planar quantum electrodynamics in a magnetic background}

\subsection{ ${\rm QED}_{2+1}$ in external homogeneous magnetic fields}

Our goal is to describe the interactions of two-dimensional electrons and
positrons with photons when there is a
constant homogeneous magnetic field in the background. We choose the
free-field Lagrangian density in the form,
\eqn
{\cal L}_0 = N \left[  c  { \bar { \psi } }(x) \left(  \gamma^{\mu} \left(  i
\hbar  \partial_{\mu} + {e\over c} A_{\mu}^{\rm ext}(x) \right)
- m c \right)  \psi(x)  -  -{ 1\over 4} f_{\mu
\nu} (x) f^{\mu \nu}(x)
- { 1\over 4} F_{\mu \nu}^{\rm ext}(x) F_{\rm ext}^{\mu \nu}(x) \right]
\label{3.1.21}
\een
after the descomposition of the electromagnetic three-vector potential in
terms of the radiation and external
fields: $A_{\mu}(x) = a_{\mu}(x) + A_{\mu}^{\rm ext}(x)$. The associated
antisymmetric tensor $F_{\mu \nu} = f_{\mu
\nu} + F_{\mu \nu}^{\rm ext}$ also splits and the quantas associated with the
field $a_{\mu}(x)$ are the planar
photons discussed in appendix A. The quanta corresponding to
$\psi(x)$ and $\bar \psi (x)$ have been
analyzed in Section \S.2.

The interaction Lagrangian density is
\eqn
{\cal L}_I = N\left[ e { \bar { \psi } } (x) \gamma^{\mu } a_{\mu }(x)
\psi(x) \right]
\label{3.1.22}
\een
and the action integral reads
\beq
 S &=& \int d^3 x \  N \left[ c  { \bar { \psi } }(x) \left( \gamma^{\mu}
\left( i
\hbar \partial_{\mu} + { e \over c} A_{\mu}(x) \right) - m c
\right)  \psi(x)
 - { 1\over 4} f_{\mu \nu}(x) f^{\mu \nu}(x) \right]  \nm \\ &-& { 1\over
4} \int d^3 x N\left[ F_{\mu \nu}^{\rm ext}(x)
F_{\rm ext}^{\mu \nu}(x) \right] \label{3.1.23}
\eeq

There is a natural length scale to the problem; the
magnetic length $l$ and the product $e^2 l =
{e^2 \over \sqrt{e B}}$ is dimensionless if $d=2$ in the n.u. system. Thus,
we express the fine structure constant
in the form:
\eqn
\alpha = {e^2 \over 4 \pi \sqrt{e B \hbar c}} \q  {\rm (c.g.s.)} \q {\rm
or} \q \alpha ={e^2 \over 4 \pi\sqrt{e B}}
\q {\rm (n.u.)} \label{3.1.23}
\een
This is consistent: in rationalized mks units the fine structure constant
is defined as
\[
\alpha = {e^2 \over 4 \pi \epsilon_0 \hbar c} \q (d=3) \q {\rm or} \q
\alpha={e^2 \over 4 \pi c_0 \sqrt{e B \hbar
c}} \q (d=2)
\]
where $c_0$ has dimensions of permitivity by length. The natural choice
$c_0=\epsilon_0 \sqrt{\hbar c \over eB}$, the
permitivity of vacuum times magnetic length, means that
\[
\alpha ={e^2 \over 4 \pi \epsilon_0 \hbar c} \equiv {e^2 \over 4 \pi c_0
\sqrt{eB \hbar c}}\approx {1\over 137.04},
\]
although the rationalized charges ${e^2 \over \epsilon_0}$ and ${e^2 \over
c_0}$ have different dimensions. We could also have defined the rationalized charge as $e^2 \over a_0$ where $a_0 = \epsilon_0 {\hbar \over m c}$, the vacuum permitivity times the electron Compton wavelength, but using the magnetic
length as length scale makes it possible to
take the limit of massless fermions in this problem.

Perturbation theory is based on the S-matrix expansion in powers of
$\alpha$. In the interaction picture the ${\rm
n}^{\rm th}$ term is the chronological product of the interaction
Hamiltonian densities at n different points,
integrated to every possible value in ${\bf R}^{1,2}$:
\eqn
S = \sum_{n=0}^{\infty} { (-i)^n \over n!} \int \cdots \int d^3 x_1 d^3 x_2
\cdots d^3 x_n \, T \left\{ {\cal H}_I(x_1){\cal H}_I(x_2) \cdots {\cal
H}_I(x_n)
\right\}  \label{3.1.24}
\een
Here, ${\cal H}_I(x) = - {\cal L}_I (x)$ and the differences with the $\vec
B =0$ case, see reference \cite{JMGMTM}, lie on the initial, final
and intermediate states in the expectation values of the S-matrix at given
orders of Perturbation Theory. From now
on we shall work in natural units.

\subsection{A process in lowest order: Compton scattering}

The transition
\[
| i \rangle = C_{n p_1}^{\dag} b^{\dag} (\vec k) | 0 \rangle \longrightarrow | f
\rangle = C_{n' {p'}_1}^{\dag} b^{\dag}(\vec{ k'}) | 0 \rangle
\]
from one electron in the $ n^{\rm th}$ Dirac-Landau level with momentum
$p_1$, $(E_n, p_1)$, and one photon
with three-momentum $k = (\omega, \vec k)$ in the initial state to one
electron in the $n'^{\rm th}$ Dirac-Landau
level with momentum $p'_1$, $(E_{n'}, p'_1)$ and one photon with $k' =(
\omega', \vec {k'})$ in the final state, is
a scattering process of amplitude $\langle f| S| i \rangle$. The dominant
contribution to this matrix element comes from  the operator
\eqn
S^{(2)} = - e^2 \int d^3 x  \, d^3 y \, \stackrel{\circ}{N} \left[   {\bar
\psi}(x) \gamma^{\alpha} a_{\alpha} (x )
i S_F(x - y) \gamma^{\beta} a_{\beta}(y) \psi(y) \right] = S_a + S_b
\label{3.2.25}
\een
which appears up to second order in $\alpha$ in the S-matrix. A subtle point is that to apply Wick's theorem we also normal-order
the creation/annihilation electron
operators in the first Landau level. This is what is meant by the symbol $
\stackrel{\circ}{N}$ and avoids tadpole photon
graphs with fermions running around the loop.

Using,
\begin{itemize}
\item  \beq
&&  \psi^+(x) | e^-; E_n,p_1 \rangle = |0\rangle {1\over \sqrt{L_1}} {\rm
u}_{n,p_1}^+ (x_2) e^{i p_1 x_1} e^{-i
E_n x_0} \nmæ\\
&& a_{\alpha}^+(x) |\gamma;\vec k \rangle = |0 \rangle {1\over \sqrt{2 A
\omega}} \epsilon_{\alpha}(\vec k) e^{-i k
x} \nm
\eeq
\item \beq
&&  \bar{\psi}^-(x) |0\rangle = \sum_{n,p_1}  | e^-; E_n,p_1 \rangle
{1\over \sqrt{L_1}} {\rm \bar u}_{n,p_1}^+
(x_2) e^{- i p_1 x_1} e^{i E_n x_0} \nmæ\\
&& a_{\alpha}^-(x) |0 \rangle =  \sum_{\vec k} |\gamma;\vec k \rangle
{1\over \sqrt{2 A \omega}}
\epsilon_{\alpha}(\vec k) e^{i k x} \nm
\eeq
\end{itemize}
one obtains
\beq
\langle f|S_a |i\rangle & =& - e^2 \int \int d^3 x d^3 y \left[ {1\over
\sqrt{L_1}} {\rm \bar u}_{n',p'_1}^+
(x_2) e^{- i p'_1 x_1+i E_{n'} x_0} \right] \left[ {1\over \sqrt{2 A \omega'}}
\gamma^{\alpha} \epsilon_{\alpha}(\vec k') e^{i k' x} \right] \nm \\
& \times & i S_F(x-y) \left[ {1\over \sqrt{2 A \omega}}
\gamma^{\beta}\epsilon_{\beta}(\vec k) e^{-i k
y} \right]  \left[ {1\over \sqrt{L_1}} {\rm u}_{n,p_1}^+ (y_2) e^{i p_1
y_1-i E_n y_0} \right]
\label{3.2.26}
\eeq
Taking into account formulas (\ref{2.1.7}), (\ref{2.1.8}) and
\[
\theta (z) = {1\over 2 \pi i} \int_{- \infty}^{\infty} d \tau { e^{i \tau
z} \over \tau - i \epsilon}, \ \lim
\epsilon \rightarrow \infty, \ \epsilon > 0
\]
the propagator can be written in the form:
\eqn
i S_F (x-y) = {l \over (2 \pi)^3 i} \int d^2 \vec q \int d \tau \left\{
\sum_{r=0}^{\infty} {e^{i q_+^{(r)}
(x-y)}\over \tau - i \epsilon} {\cal U}_{r q_1}^+ (q_2)  {\bar {\cal U}}_{r
q_1}^+ (q_2) -
\sum_{r=1}^{\infty} {e^{i q_-^{(r)} (x-y)}\over \tau - i \epsilon} {\cal
U}_{r q_1}^- (q_2)  {\bar {\cal U}}_{r
q_1}^- (q_2)  \right\} \label{3.2.27}
\een
where $q_{\pm}^{(r)} = ( \pm q_0^{(r)}, -\vec q)$ with $q_0^{(r)} = \tau -
E_r$ and $\vec q = (q_1, q_2)$.
Also plugging the Fourier transform of the ongoing and
outgoing spinors in (\ref{3.2.26}),  we are ready to perform
the x- and y-integrations. The outcome of the $x_0$- and $y_0$-integrations
is energy conservation:
\beq 
 && \int d x_0 \ {\rm exp} [ i x_0 \ ( E_{n'} + \omega' \mp E_r \pm \tau )]
\int d y_0 \ {\rm exp} [ i y_0\  (
\pm E_{r} \mp \tau - E_n - \omega )] \nm \\
&& = (2 \pi)^2 \delta ( E_{n'} + \omega' \mp E_r \pm \tau ) \delta ( \pm
E_{r} \mp
\tau - E_n - \omega ) \nm
\eeq
Simili modo, the spatial integrations
\beq
&& \int d x_j  \ {\rm exp} [ i x_j  \ ( q_j - p'_j - k'_j)] \int d y_j \
{\rm exp} [ i y_j\  ( p_j +
k_j - q_j )] \nm \\
&& =  (2 \pi)^2 \delta ( q_j - p'_j - k'_j) \delta( p_j +
k_j - q_j ), \nm
\eeq
lead to momentum conservation.

The $\delta$-functions allow us to compute the $\tau$-, $q_1$- and
$q_2$-integrations. We obtain:
\eqn
\langle f| S_a | i \rangle =2 \pi l^2 \int \int d p_2 d p'_2 \left[ \delta
(E_{n'} +
\omega' - E_n -\omega) \delta^{(2)} ({\vec p'} + {\vec k'} - \vec p -\vec k
) {1\over L_1} \cdot {1\over 2 A
\omega} \right] {\cal M}_a (p_2, p'_2) \label{3.2.28}
\een
where
\eqn
{\cal M}_a (p_2, p'_2) = -e^2 {\bar {\cal U}}_{n' p'_1}^+ (p'_2)
\gamma^{\alpha} \epsilon_{\alpha}
(\vec{k'}) i S_F (E_n +\omega, \vec p  + \vec k )  \gamma^{\beta}
\epsilon_{\beta} (\vec k)  {\cal
U}_{n p_1}^+ (p_2) \label{3.2.29}
\een
is the Feynman. The momentum
space propagator reads:
\beq
&& i S_F (E_n +\omega, \vec p  + \vec k) \label{3.2.30} \\
&& = {l\over i} \left\{ \sum_{r = 0}^{\infty} {{\cal
U}_{r,(p_1+k_1)}^+ (p_2+k_2) {\bar {\cal U}}_{r,(p_1+k_1)}^+ (p_2+k_2)
\over E_r - (E_n + \omega ) - i \epsilon
} - \sum_{r=1}^{\infty}{ {\cal U}_{r,(p_1+k_1)}^- (p_2+k_2) {\bar {\cal
U}}_{r,(p_1+k_1)}^- (p_2+k_2) \over E_r +
(E_n +
\omega ) - i \epsilon } \right\} \nm
\eeq

The matrix element $\langle f| S_b | i \rangle$ corresponding to the
exchange graph is given by an analogous
expression to (\ref{3.2.28}),  ${\cal M}_a (p_2,p'_2)$ being replaced by the
Feynman amplitude:
\eqn
{\cal M}_b (p_2, p'_2) = -e^2 {\bar {\cal U}}_{n' p'_1}^+ (p'_2)
\gamma^{\alpha} \epsilon_{\alpha} ({\vec
k}) i S_F (E_n - \omega', \vec p - {\vec k'})  \gamma^{\beta}
\epsilon_{\beta} (\vec{ k'}) {\cal
U}_{n p_1}^+ (p_2) \label{3.2.31}
\een
\beq
&& i S_F (E_n -\omega', \vec p- {\vec k'}) \label{3.2.32} \\
&& = {l\over i} \left\{ \sum_{r =0}^{\infty}  { {\cal U}_{r,(p_1-k'_1)}^+
(p_2-k'_2) {\bar {\cal
U}}_{r,(p_1-k'_1)}^+ (p_2-k'_2) \over E_r - (E_n - \omega' ) - i \epsilon }
- \sum_{r=1}^{\infty} { {\cal U}_{r,
(p_1-k'_1)}^- (p_2-k'_2) {\bar {\cal U}}_{r,(p_1-k'_1)}^- (p_2-k'_2) \over
E_r + (E_n -\omega' ) - i \epsilon }
\right\} \nm
\eeq

We express the result for the S-matrix element in the form
\[
 \langle f | S^{(2)} | i \rangle \equiv S_{fi}^{(2)}= {l^2 \over (2 \pi )}
\int d p_2 \, S_{fi}^{(2)} (p_2,
p_2+k_2-k'_2)
\]
\beq
&& S_{fi}^{(2)} (p_2, p_2+k_2-k'_2) = (2 \pi )^2 \delta (E_{n'} + \omega' -
E_n - \omega)
\delta({p'}_1 + {k'}_1 - p_1 - k_1 )  \prod_{\rm ext} \left( {1\over L_1}
\right) ^{1/2}  \nm \\ && \times
\prod_{\rm ext} \left( {1\over 2 A \omega_{\vec k} } \right) ^{1/2}
 ({\cal M}_a(p_2, p_2+k_2-k'_2) + {\cal M}_b(p_2, p_2+k_2-k'_2)) \label{3.2.33}
\eeq
Note that despite the formal identity with the scattering amplitudes for
the planar Compton effect at zero
external magnetic field, there are profound differences: (1) The
Dirac-Landau spinors and the fermion propagator
include Hermite polynomials that depends on the momentum in the
$OX_2$-direction. (2) Because there is no invariance
with respect to translations in the $OX_2$-direction, the initial and final
states are not eigenvalues of the $\hat
p_2$ operator; thus, we obtain contributions from all the possible eigenvalues
$p_2$ and $p'_2$ and we need
to integrate over their full range. Nevertheless, there is invariance
under magnetic traslations, see \cite{Capelli93}, and
because of this $p_2$ and $p'_2$ are not completely independent but
related by the condition: $p'_2 = p_2 + k_2
- k'_2$.

\subsection{Feynman rules for ${\rm QED}_{2+1}$ in a magnetic field}

Following the pattern shown in the derivation of the planar Compton effect,
it is possible to establish a set of
Feynman rules for writing the S-matrix elements directly for the Feynman
graphs in ${\rm QED}_{2+1}$ when there is
an external magnetic field such that $\vec B (x) = - B \vec k$.

The initial and final states for any process are tensor products of photons
occupying plane wave states and fermions
in Dirac-Landau states. Thus, the \lq\lq quantum" numbers are the photon
momenta and the energies and momenta in
the $OX_1$-direction of the fermions. For the transition $|i\rangle
\rightarrow |f\rangle$, the S-matrix element
takes the form:
\eqn
 \langle f | S| i \rangle = {l \over 2 \pi} \int d p_2^{(1)} {l \over 2
\pi} \int d p_2^{(2)} \cdots
{l\over 2 \pi} \int d p_2^{(n-1)}  l  S_{fi}(p_2^{(1)}, p_2^{(2)}, \cdots,
p_2^{(n-1)}, P_2^i - \stackrel{\circ}{
P}_2^f ) \label{3.3.34}
\een
where $p_2^{(1)}, p_2^{(2)}, \cdots, p_2^{(n)}$ are the momenta in the
$OX_2$-direction of the external fermions.
Here, we have:
\beq
P_1^i &=& \sum_{a=1}^{n-m} p_1^{i(a)} + \sum_{a=1}^{N-M} k_1^{i(a)} \ , \
P_2^i = \sum_{a=1}^{n-m} p_2^{i(a)} +
\sum_{a=1}^{N-M} k_2^{i(a)} \nm \\
P_1^f &=& \sum_{a=1}^{m} p_1^{f(a)} + \sum_{a=1}^{M} k_1^{f(a)} \ , \
\stackrel{\circ}{P}_2^f = \sum_{a=1}^{m-1}
p_2^{f(a)} + \sum_{a=1}^{M} k_2^{f(a)} \label{3.3.35}
\eeq
\[
p_2^{(n)} = p_2^{f(m)} = P_2^i - \stackrel{\circ}{P}_2^f \ ,
\]
and,
\beq
 S_{fi}(p_2^{(1)}, p_2^{(2)}, \cdots, p_2^{(n-1)}, P_2^i -
\stackrel{\circ}{ P}_2^f )&&= \delta _{fi} + \left[ (2\pi
)^2 \delta ( E_f - E_i) \delta (P_1^f - P_1^i)  \right. \label{3.3.36} \\
&& \left. {\d \prod_{\rm ext}}
\left( {1 \over \sqrt{L_1}}\right)   {\d \prod_{\rm ext}} \left( {1 \over
\sqrt{2 A \omega}} \right)   \right]
{\cal M} (p_2^{(1)}, p_2^{(2)}, \cdots, p_2^{(n-1)}, P_2^i - \stackrel{\circ}{ P}_2^f ) \nm
\eeq
$E_i$ and $E_f$ are the total energies of the initial and final states; the
products extend over all external
fermions and photons with normalization factors $1/\sqrt{L_1}$ and
$1/\sqrt{2 A \omega}$, respectively.

The Feynman amplitude ${\cal M} (p_2^{(1)}, p_2^{(2)}, \cdots, p_2^{(n-1)},
P_2^i - \stackrel{\circ}{ P}_2^f )$ is
the sum of the contributions ${\cal M}^{(m)} (p_2^{(1)}, p_2^{(2)}, \cdots,
p_2^{(n-1)}, P_2^i - \stackrel{\circ}{
P}_2^f )$ for all orders in perturbation theory. The contribution to ${\cal
M}^{(m)}$ from each topologically
different graph is obtained from the following Feynman rules:
\begin{itemize}
\item For each vertex, write a factor $i e \gamma^{\alpha}$.
\item For each internal photon line, labelled by the three-momentum $k$,
write a factor: $ i D_{F\alpha \beta} (k) =
i {- g_{\alpha \beta} \over k^2 + i \epsilon}$.
\item For each internal fermion line labelled by the energy $E$ and the
momentum $\vec q = (q_1, q_2)$, write a
factor
\eqn
i S_F(E, \vec q ) = {l\over i} \left[ \sum_{r=0}^{\infty}  {
{\cal U}_{r,q_1}^{+}(q_2)  {\bar {\cal U}}_{r,q_1}^{+}(q_2) \over E_r - E - i
\epsilon } - \sum_{r=1}^{\infty} { {\cal U}_{r,q_1}^{-}(q_2)  {\bar {\cal
U}}_{r,q_1}^{-}(q_2) \over E_r + E - i
\epsilon }   \right]  \label{3.3.37}
\een
\item For each external line, write one of the following factors:
\begin{itemize}
\item for each initial electron: ${\cal U}_{n,p_1}^+(p_2)$ or ${\cal
U}_{0,p_1}^+ (p_2)$
\item for each initial positron: ${\cal U}_{n,p_1}^-(p_2)$
\item for each final electron: ${\bar {\cal U}}_{n,p_1}^+(p_2)$ or ${\bar
{\cal U}}_{0,p_1}^0(p_2)$
\item for each final positron: ${\bar {\cal U}}_{n,p_1}^-(p_2)$
\item for each initial or final photon: $\epsilon_{1 \alpha} (\vec k)$
\end{itemize}
\item There is a ${l\over 2 \pi} \int d p_2$ integration for each external
fermion.
\item Because of energy conservation, $E = \sum_a E_a $, where $E_a$ is the
energy of each particle created or
annihilated at any given vertex. There is also momentum conservation at
each vertex.
\item For each photon three-momentum that is not fixed by energy-momentum
conservation carry out the integration
${1\over (2 \pi)^3 }\int d^3 q$. One such integration with respect to an
internal photon momentum occurs for each
closed loop.

\item For each closed fermion loop, there is one fermion energy and one
momentum which are not fixed by energy-momentum conservation. One must
perform the integration over these variables and also take the trace and
multiply by a factor $(-1)$. For instance, for the graph of vacuum
polarization by
Dirac-Landau electrons one obtains:

\beq
\Pi_{\alpha \beta}[k] &=& {e^2 \over (2\pi)^3} \int d^2 \vec p \int d \lambda
\ {\rm Tr} \left[ \sum_{m=0}^{\infty} i S_F (E_m - \lambda -\omega,
\vec p - \vec k) \gamma_{\alpha} i S_{F_m}^+ (E_m-\lambda, \vec p) \gamma_{\beta} \right. \\
&-& \left. \sum_{m=1}^{\infty} i S_F (- E_m + \lambda -\omega,
\vec p - \vec k) \gamma_{\alpha} i S_{F_m}^- (- E_m + \lambda, \vec p) \gamma_{\beta} \right]
\eeq

where $ i S_F (\pm E_m \mp \lambda -\omega, \vec p - \vec k )$ is given by (\ref{3.3.37}) with $ E = \pm E_m \mp \lambda -\omega $ and $\vec q = \vec p - \vec k $ and 

\[
i S_{F_m}^{\pm} (\pm E_m \mp \lambda, \vec p) = {l\over i}\cdot {
{\cal U}_{m,p_1}^{\pm}(p_2)  {\bar {\cal U}}_{m,p_1}^{\pm}(p_2) \over \lambda - i \epsilon }
\]
\end{itemize}

\section{ Compton effect on Dirac-Landau electrons}

In this Section we shall discuss the scattering cross-length for the
Compton effect on electrons in a constant
magnetic field up to second order in perturbation theory. We bear in mind
the quantum Hall effect where a
two-dimensional gas of electrons at very low temperatures is subjected to a
strong magnetic field, from 6 to 26
Teslas, which is constant, uniform and perpendicular to the plane where the
electrons move.

\subsection{ The scattering amplitude of photons by Dirac-Landau electrons}

Under these circumstances we expect the electrons to be in the lowest (non-
filled) Landau level. Thus, we consider
the initial and final electrons occupying two states in the first Landau
level. The S-matrix element and the
Feynman amplitude for the transition $|i\rangle = | e^-(E_0, p_1);
\gamma(k) \rangle \rightarrow |f\rangle = | e^-(
E_0, p'_1); \gamma(k') \rangle $ are given by:
\beq
\langle f | S | i \rangle &=& { \pi l^2 \delta(\omega'-\omega) \delta
(p'_1+k'_1-p_1-k_1) \over A L_1
\sqrt{\omega \omega'}} \int\int d p_2 d p'_2 \delta(p'_2+k'_2-p_2-k_2)  \nm \\ &&
\q \q \q \q \q\q\q \times \left[ {\cal M}_a(p_2,p'_2)+{\cal
M}_b(p_2,p'_2)\right]  \nm \\
 {\cal M}_a (p_2, p'_2) &=&  i e^2 \omega l \epsilon (\vec k') \bar
\epsilon (\vec k) \varphi_{0 p'_1}^* (p'_2)
\varphi_{0 p_1} (p_2) \sum_{r = 1}^{\infty} { | \varphi_{r-1, p_1+ k_1}
(p_2 + k_2)|^2 \over E_r^2 -( m + \omega)^2
} \label{4.1.38} \\
 {\cal M}_b (p_2, p'_2) &=& - i e^2 \omega l \epsilon (\vec k) \bar
\epsilon (\vec k') \varphi_{0 p'_1}^* (p'_2)
\varphi_{0 p_1} (p_2) \sum_{r = 1}^{\infty}   { | \varphi_{r-1, p_1- k'_1}
(p_2 - k'_2)|^2 \over E_r^2 -( m -
\omega')^2 }   \nm
\eeq

To obtain these formulas we have re-written (\ref{3.2.29}) and
(\ref{3.2.31}) according to the following
information:
\begin{itemize}
\item \[ \bar {\cal U}_{0 p'_1}^+ (p'_2) = \left( \ba{cc} \varphi_{0
p'_1}^* (p'_2) & 0 \ea \right)  \q , \q {\cal
U}_{0 p_1}^+ (p_2) = \left( \ba{c} \varphi_{o p_1}(p_2) \\ 0 \ea \right)
\]
\item $\epsilon_{\alpha}^{(1)} (\vec k) = ( 0, {\vec \epsilon }^{(1)} (\vec
k))$ such that ${\vec \epsilon}^{(1)}
(\vec k) \cdot \vec k = 0$ is the only transversal polarization vector of
the incoming planar photon. We define
$\epsilon (\vec k) = \epsilon_2^{(1)}(\vec k) + i \epsilon_1^{(1)}(\vec k)
$ and its complex conjugate: $\bar \epsilon (\vec k) =
\epsilon_2^{(1)}(\vec k) - i \epsilon_1^{(1)}(\vec k) $. Then,
\[
\epsilon_{\alpha}^{(1)} (\vec k) \gamma^{\alpha} = \left( \ba{cc} 0 &
\epsilon (\vec k) \\ - \bar \epsilon (\vec k)
& 0 \ea \right)
\]
There is an identical formula for the outgoing photon.
\item The fermion propagator splits into three parts:
\[
i S_F (E,q_1, q_2) = i \left[ S_F^+ (E, q_1, q_2)-  S_F^- (E, q_1, q_2 ) +
S_F^0
(E,q_1, q_2) \right]
\]
where
\beq
S_F^{\pm} (E, q_1, q_2) &=& - l  \sum_{r = 1}^{\infty} {E_r \pm m   \over 2
E_r (E_r \mp E - i
\epsilon )} \nm \\ && \times  \left( \ba{cc} \varphi_{rq_1}(q_2)
\varphi_{rq_1}^*(q_2)  & \pm{E_r^0 \over E_r \pm
m} \varphi_{rq_1}(q_2)  \varphi_{r-1 q_1}^*(q_2) \\ \mp {E_r^0 \over E_r
\pm m} \varphi_{r-1 q_1}(q_2)  \varphi_{r
q_1}^*(q_2) & - \left( {E_r^0 \over E_r \pm m}\right) ^2 \varphi_{r-1
q_1}(q_2)  \varphi_{r-1
q_1}^*(q_2) \ea \right) \nm
\eeq
and
\[
S_F^0 (E, q_1, q_2) = - {l \over E_0 - E - i \epsilon }\left( \ba{cc}
\varphi_{0q_1}(q_2)
\varphi_{0q_1}^*(q_2)  & 0 \\ 0 & 0 \ea \right)
\]
In our process we have $E = m + \omega$, $q_1 = p_1 + k_1$ and
$q_2=p_2+k_2$ for ${\cal M}_a(p_2, p'_2)$ and $E= m
- \omega'$, $q_1= p_1- k'_1$ and $q_2 = p_2 - k'_2$ for ${\cal M}_b (p_2,
p'_2)$; the factor $E_r^0 = +
\sqrt{E_r^2 - m^2}$ appears in the normalization.
\end{itemize}

Now replacing,
\begin{itemize}
\item $ \varphi_{0p'_1}^*(p'_2)  \varphi_{0p_1}(p_2) = {2 \sqrt{\pi} \over
l} {\rm exp} \left[ -i (p'_1 p'_2-
p_1 p_2) l^2\right]  {\rm exp} \left[ -{1\over 2} (p_2^2 + {p'_2}^2 ) l^2
\right]
$
\item $ E_r^2 -(m+\omega)^2 = 2 e Bæ[r + c] \ , \ c=- {\omega (\omega +2 m)
\over 2 e B} $
\item $ E_r^2 -(m-\omega')^2 = 2 e Bæ[r + c'] \ , \ c'=- {\omega' (\omega'
- 2 m) \over 2 e B} $
\item $ |\varphi_{r-1,q_1}(q_2)|^2 ={2 \sqrt{\pi} \over l} \cdot {1\over
2^{(r-1)} (r-1)!} H_{r-1}^2 [q_2 l]\ e^{  -q_2^2 l^2} $
\end{itemize}
in ${\cal M}_a$, where $q_1 = p_1+k_1$, $q_2=p_2+k_2$, and in ${\cal M}_b$,
where $q_1=p_1-k'_1$, $q_2=p_2-k'_2$, we
obtain:
\beq
{\cal M}_a (p_2, p'_2) &=&  { i 2 \pi \sqrt{\pi} \over l B} e \omega
\epsilon (\vec k') \bar \epsilon (\vec k)
{\rm exp} \left[ - i (p'_1p'_2-p_1p_2) l^2 \right] {\rm exp} \left[
-{1\over 2} (p_2^2+{p'_2}^2) l^2 \right] \nm \\
&& \q\q\q\q\q \times  \sum_{r = 1}^{\infty} { H_{r-1}^2[(p_2+k_2) l] \over
\sqrt{\pi} 2^{(r-1)} (r-1)!} \cdot {
e^{-(p_2+k_2)^2 l^2} \over r + c} \label{4.1.39} \\
{\cal M}_b (p_2, p'_2) &=&  {- i 2 \pi \sqrt{\pi} \over l B} e \omega
\epsilon (\vec k) \bar \epsilon (\vec k')
{\rm exp} \left[ - i (p'_1p'_2-p_1p_2) l^2 \right] {\rm exp} \left[
-{1\over 2} (p_2^2+{p'_2}^2) l^2 \right] \nm \\
&& \q\q\q\q\q \times  \sum_{r = 1}^{\infty} { H_{r-1}^2[(p_2-k'_2) l] \over
\sqrt{\pi} 2^{(r-1)} (r-1)!} \cdot {
e^{-(p_2-k'_2)^2 l^2} \over r + c'} \nm
\eeq

In order to sum the series in (\ref{4.1.39}), we consider the spectral problem
\[
 \Delta_c \psi_n(x) = \lambda_n^c \psi_n(x)
\]
\[ \lambda_n^c = n + c + 1 \ , \ n = 0, 1, 2, \cdots \q ; \q \psi_n(x) =
{1\over \pi^{1/4} \sqrt{ 2^n n!}} H_n[x]
e^{-{1\over 2} x^2} \ ,\]
for the elliptic differential operator $ \Delta_c = {1\over 2} \left[ -
{d^2 \over d x^2} + x^2 + 2 c +1 \right] $
on $L^2({\rm {\bf R}})$. The Green function for the s-power of $\Delta_c$
is defined as:
\[
G_{\Delta_c} (x,y;s) = \sum_{n=0}^{\infty} { \psi_n^*(x) \psi_n(y) \over
(\lambda_n^c)^s}
\]
We immediately notice that:
\beq
G_{\Delta_c}(x,x;1) &=& \sum_{r = 1}^{\infty} { H_{r-1}^2[(p_2+k_2) l]
\over \sqrt{\pi} 2^{(r-1)} (r-1)!} \cdot {
e^{-(p_2+k_2)^2 l^2} \over r + c}  \ , \ x = (p_2+k_2) l \nm \\ &&
\label{4.1.40} \\
G_{\Delta_{c'}}(x,x;1) &=& \sum_{r = 1}^{\infty} { H_{r-1}^2[(p_2-k'_2) l]
\over \sqrt{\pi} 2^{(r-1)} (r-1)!} \cdot {
e^{-(p_2-k'_2)^2 l^2} \over r + c'}  \ , \ x = (p_2-k'_2) l \nm
\eeq
This is related to the heat kernel
\[
K_{\Delta_c}(x,y;\beta) = \sum_{n=0}^{\infty} e^{-\beta \lambda_n^c}
\psi_n^*(x) \psi_n(y)
\]
of the operator $\Delta_c$  through a Mellin transform:
\[
G_{\Delta_c}(x,y;s) = {1\over \Gamma[s]} \int_0^{\infty} d \beta
K_{\Delta_c} (x,y;\beta)
\]
We write $\Delta_c = \Delta +c + 1/2$, so that,
\[
K_{\Delta_c}(x,y;\beta) = e^{- \beta(c+{1\over 2})} K_{\Delta} (x,y;\beta)
\]
The heat kernel for the differential operator $\Delta $ of the Harmonic
oscillator is very well known
\cite{Feynman} and we obtain:
\beq
G_{\Delta_c}(x,x;1) &=& {1\over \Gamma[1]} \int_0^{\infty} d \beta
K_{\Delta_c} (x,x;\beta) \nm\\ && \label{4.1.41}
\\ K_{\Delta_c}(x,x;\beta) &=& { e^{-\beta(c+{1\over 2})} \over \sqrt{2\pi
{\rm sinh} \beta}} {\rm exp} \left[ - x^2
{\rm tanh} {\beta\over 2} \right] , \nm
\eeq
and a similar expression for $G_{\Delta_{c'}}(x,x;1)$. Before performing
the integration we plug the integral form
of $G_{\Delta_c}$ and $G_{\Delta_{c'}}$ into
\[
{\cal M} = \int dp_2 \int dp'_2 \delta(p'_2+k'_2-p_2-k_2) \left[ {\cal
M}_a(p_2,p'_2) +{\cal M}_b(p_2,p'_2) \right]
.
\]
We first integrate in the variables $p_2$ and $p'_2$ and then in $\beta$ to
reach the finite answer:
\beq
&&{\cal M}_a  =  { i \sqrt{2} \pi^2  \over l^2 B} e \omega  \epsilon (\vec
k') \bar \epsilon (\vec k)
{\rm exp} \left[ i (p'_1k'_2-p_1k_2)^2 l^2 - {1\over 2} (k_2^2+{k'_2}^2)
l^2 \right] \nm \\ && {\rm exp}
\left[ - {[(p'_1-p_1) + i (k_2+k'_2) ]^2 l^2 \over 8} \right]  {\Gamma[c+1]
\over
\Gamma[c+3/2]} \ _1 F_1 \left[ c+1,c+ {3\over 2}; {[(k_2+k'_2)- i
(p'_1-p_1) ]^2 l^2 \over 8}\right] \nm \\
&&  \label{4.1.42} \\
&& {\cal M}_b  =  {- i \sqrt{2} \pi^2  \over l^2 B} e \omega  \epsilon
(\vec k) \bar \epsilon (\vec k')
{\rm exp} \left[ - i (p'_1k_2-p_1k'_2)^2 l^2 - {1\over 2} (k_2^2+{k'_2}^2)
l^2 \right] \nm \\
&&  {\rm exp} \left[ -
{[(p'_1-p_1) - i (k_2+k'_2) ]^2 l^2 \over 8} \right]   {\Gamma[c'+1] \over
\Gamma[c'+3/2]} \ _1 F_1 \left[ c'+1,c'+ {3\over 2}; {[(k_2+k'_2)+ i
(p'_1-p_1) ]^2 l^2
\over 8}\right] \nm
\eeq
where $\ _1F_1[a,b;z]$ is a degenerated confluent hypergeometric function,
see \cite{Gradshteyn}. The integration in the
variable $p'_2$ is performed immediately because of the $\delta$-function; we
are left with Gaussian integrals in $p_2$ which
can be easily calculated. Finally, the integration in the $\beta$ variable
produces a degenerated hypergeometric
function. If we had chosen to integrate first in the $\beta$ variable, we
would have used some regularization
procedure to avoid the dangerous singularity at $\beta =0$ that renders
$G_{\Delta_c}(x,x;1)$ strictly divergent.

\subsection{ The differential cross-length}

The differential  cross-length is the transition rate into a group of final
states for a scattering centre
and unit incident flux. If $\omega = {|S_{fi}|^2 \over T}$ is the
transition probability per unit of time for our
process, we have:
\[
d \lambda = \omega {1\over v_{\rm rel} / A}   \cdot  {d p'_1 \over
(2\pi)}\cdot  {d^2 \vec k' \over (2\pi)^2 }
\]
where we use finite normalization time $T$ and area $A$. There are $ { A
d^2 \vec k' \over (2 \pi)^2} $ photon
final states with momentum belonging to the interval $(\vec k', \vec k' + d
\vec k')$ and ${L_1 d p'_1 \over (2
\pi)}$  electron final states in the first Landau level and $p_1^f \in
(p'_1, p'_1 + d p'_1)$. $v_{\rm rel} /A$ is
the incident flux of incoming particles. Using
\[
\left[ \delta (\omega - \omega') \delta (p'_1+ k'_1 - p_1 - k_1) \right] ^2
= {T L_1 \over (2\pi )^2 } \delta
(\omega - \omega') \delta (p'_1+ k'_1 - p_1 - k_1)
\]
we obtain
\[
d \lambda = (2\pi)^2 \delta (\omega' - \omega) \delta (p'_1+ k'_1 - p_1 -
k_1)\ {l^4 \over 16 \pi^2 \omega v_{\rm
rel}} \ {d p'_1 \over (2\pi)}\  {d^2 \vec k' \over (2\pi)^2 2 \omega'}\
|{\cal M} |^2
\]

Because of the conservation of energy and momentum it is easy to integrate this
expression with respect to $p'_1$ and
$|\vec k'|$. Note that $d^2 \vec k' =|\vec k'| d|\vec k'| d \theta $, and
 \[
{ d \lambda \over d \theta} = { l^4 \over 32 \pi^2  \omega v_{\rm rel} }
|{\cal M} |^2
\]
bearing in mind that $p'_1 = p_1 + k_1 - k'_1$ and $ \omega' = \omega$.

The above result is referred to a general reference frame. We choose the
laboratory frame characterized by:

1) \underline {initial state}; one electron in the lowest Dirac-Landau
state with energy $E_0=m$ and momentum
$p_1=0$ plus a photon of momentum $k=(\omega, \vec k)$, $\vec k = (k_1, 0)$
and polarization ${\vec
\epsilon}^{(1)}( \vec k) = (0, 1)$.

2) \underline {final state}; one electron also in the lowest Dirac-Landau
state but with momentum $p'_1 \neq 0$
plus a photon with momentum $k'=(\omega', \vec k')$, such that $\vec k
\cdot \vec k' = \omega \omega' \cos \theta$, and
polarization ${\vec \epsilon}^{(1)}( \vec k') = (-\sin \theta, \cos
\theta)$.

Conservation of energy and momentum requires that,
\beq
&& E_0 + \omega = E_0 + \omega' \ , \  \omega ' = \omega \nm \\
&& p_1 + k_1 = p'_1 + k'_1 \ , \  p'_1 = \omega (1- \cos \theta ) \nm
\eeq
There is no Compton shift in wavelength for the photon because the energies
of the final and initial electrons are
the same; however, there is a shift in the electron momentum. The recoil
angle of
the electron is given by $\cos \beta_r
= {\vec p' \vec k \over |\vec p'| |\vec k|} = { p'_1 \over |\vec p'|}$, and
the relative velocity is $v_{\rm rel} =
|\vec k| / \omega = 1$. Inelastic scattering would require a different
Dirac-Landau state for the outgoing electron.

In the laboratory frame the calculation of $|{\cal M}|^2 = {\cal M}_a {\cal
M}_a^* + {\cal M}_a {\cal M}_b^* +{\cal
M}_b {\cal M}_a^* + {\cal M}_b {\cal M}_b^* $ $ = X_{aa} + X_{ab} + X_{ba}
+ X_{bb}$, gives:
\beq
 X_{aa} &=& {2 \pi^4 e^2 \over l^4} {\omega^2 \over B^2} {\rm exp} \left[
-{\omega^2 l^2 \over 2}  \left( \sin^2
\theta + 2 \sin^2 {\theta \over 2} \right) \right] \left[ {\Gamma[c+1]
\over \Gamma[c+3/2]} \right] ^2 \nm \\
& \times & \left| \ _1 F_1 \left[ c+1,c+ {3\over 2}; {\omega^2 l^2 \over 2}
\left( {\sin \theta - i (1 - \cos
\theta ) \over 2} \right) ^2 \right] \right| ^2 \nm \\
 X_{bb} &=& {2 \pi^4 e^2 \over l^4} {\omega^2 \over B^2} {\rm exp} \left[
-{\omega^2 l^2 \over 2}   \left( \sin^2
\theta + 2 \sin^2 {\theta \over 2} \right) \right] \left[ {\Gamma[c'+1]
\over \Gamma[c'+3/2] } \right] ^2 \nm \\
&  \times & \left| \ _1 F_1 \left[ c'+1,c'+ {3\over 2}; {\omega^2 l^2 \over
2} \left( {\sin \theta + i (1 - \cos
\theta ) \over 2} \right) ^2 \right] \right| ^2 \nm
\eeq
\beq
&& X_{ab} = X_{ba}^* = - {2 \pi^4 e^2 \over l^4} {\omega^2 \over B^2} {\rm
exp} \left[ -{\omega^2 l^2 \over 2}
\left( \sin^2 \theta + 2 \sin^2 {\theta \over 2} \right) \right]
{\Gamma[c+1] \over \Gamma[c+3/2] }{\Gamma[c'+1]
\over \Gamma[c'+3/2] }
\nm \\ && \times   {\rm exp} \left[ - 2 i \theta + i \omega^2 l^2 \sin
\theta \sin^2 {\theta \over  2} \right] \ _1
F_1 \left[ c+1,c+ {3\over 2}; {\omega^2 l^2 \over 2} \left( {\sin \theta -
i (1 - \cos
\theta ) \over 2} \right) ^2 \right] \nm \\
&&  \times \ _1 F_1 \left[ c'+1,c'+ {3\over 2}; {\omega^2 l^2 \over 2}
\left( {\sin
\theta + i (1 - \cos \theta ) \over 2} \right) ^2 \right]
 \nm
\eeq
We finally write the differential cross-length for the scattering of
photons by Dirac-Landau electrons in the plane:
\beq
&& \left( {d \lambda \over d \theta} \right) _{\rm Lab} = { \alpha^2 \pi^3
\omega \over e B} {\rm exp} \left[
-{\omega^2 l^2 \over 2}  \left( \sin^2
\theta + 2 \sin^2 {\theta \over 2} \right) \right] \times \nm \\
&& \left\{ \left[ {\Gamma[c+1] \over \Gamma[c+3/2]} \right] ^2  \left| \ _1
F_1 \left[ c+1,c+ {3\over 2};
{\omega^2 l^2 \over 2} \left( {\sin \theta - i (1 - \cos
\theta ) \over 2} \right) ^2 \right] \right| ^2  +  \left[ {\Gamma[c'+1]
\over \Gamma[c'+3/2] } \right] ^2
\right. \nm \\
&& \left| \ _1 F_1 \left[ c'+1,c'+ {3\over 2}; {\omega^2 l^2 \over 2}
\left( {\sin \theta + i (1 -
\cos \theta ) \over 2} \right) ^2 \right] \right| ^2 -  2 {\Gamma[c+1]
\over \Gamma[c+3/2] }{\Gamma[c'+1]
\over \Gamma[c'+3/2] } \nm \\
&& {\rm Re} \left[  {\rm exp} \left[ - 2 i \theta + i \omega^2 l^2 \sin
\theta \sin^2 {\theta \over  2} \right] \ _1
F_1 \left[ c+1,c+ {3\over 2}; {\omega^2 l^2 \over 2} \left( {\sin \theta -
i (1 - \cos
\theta ) \over 2} \right) ^2 \right]  \right. \nm \\
&& \left. \left. \times \ _1 F_1 \left[ c'+1,c'+ {3\over 2}; {\omega^2 l^2
\over 2} \left( {\sin
\theta + i (1 - \cos \theta ) \over 2} \right) ^2 \right] \right] \right\}
\label{4.2.43}
\eeq

\subsection{Angular distribution and total cross-length }

In this Section we shall discuss the physical meaning of the important
formula (\ref{4.2.43}). It is convenient to
express the differential cross-length in terms of the dimensionless
constants $\gamma= {\omega \over  m}$ and
$\beta = {\sqrt{2 e B} \over m}$, and also to introduce a new constant,
$L_T = {2 \pi^2 \alpha^2 \over \sqrt{2 e
B}}$, which is a length associated with the system. Equation (\ref{4.2.43})
becomes:
\beq
&&{1\over \pi L_T} \left( {d \lambda \over d \theta} \right) _{\rm Lab} = {
\gamma \over \beta} {\rm
exp} \left[ -\left( {\gamma \over \beta} \right) ^2  \left( \sin^2 \theta +
2 \sin^2 {\theta \over 2} \right)
\right] \times \nm \\
&& \left\{ \left[ {\Gamma[1- {\gamma (\gamma +2) \over \beta^2}] \over
\Gamma[3/2- {\gamma (\gamma +2) \over
\beta^2}]} \right] ^2  \left| \ _1 F_1
\left[ 1- {\gamma (\gamma +2) \over \beta^2}, {3\over 2}- {\gamma (\gamma
+2) \over \beta^2}; \left( {\gamma
\over \beta} \right) ^2 \left( {\sin \theta - i (1 - \cos \theta ) \over 2}
\right) ^2 \right] \right| ^2  +
\right. \nm \\
&&  \left[ {\Gamma[1- {\gamma (\gamma -2) \over \beta^2}] \over \Gamma[3/2-
{\gamma (\gamma -2) \over \beta^2}] }
\right] ^2 \left| \ _1 F_1 \left[ 1- {\gamma (\gamma -2) \over
\beta^2},{3\over 2}- {\gamma (\gamma -2) \over
\beta^2}; \left( {\gamma \over \beta} \right) ^2 \left( {\sin \theta + i (1
- \cos \theta ) \over 2} \right) ^2
\right] \right| ^2 - \nm \\
&&  2 {\Gamma[1- {\gamma (\gamma +2) \over \beta^2}] \over \Gamma[3/2-
{\gamma (\gamma +2) \over \beta^2}]
}{\Gamma[1- {\gamma (\gamma -2) \over \beta^2}] \over \Gamma[3/2- {\gamma
(\gamma -2) \over \beta^2}] } {\rm Re}
\left[  {\rm exp} \left[ - 2 i \theta + 2 i \left( {\gamma \over \beta}
\right) ^2 \sin \theta \sin^2 {\theta
\over  2} \right]  \right. \nm \\
&& \q  \times \ _1 F_1 \left[ 1- {\gamma (\gamma +2) \over \beta^2},
{3\over 2}- {\gamma (\gamma +2) \over
\beta^2};
\left( {\gamma \over \beta} \right) ^2 \left( {\sin \theta - i (1 - \cos
\theta ) \over 2} \right) ^2 \right]
\nm \\
&& \left. \left. \q  \times   \ _1 F_1 \left[ 1- {\gamma (\gamma -2) \over
\beta^2}, {3\over 2}-
{\gamma (\gamma -2) \over \beta^2}; \left( {\gamma \over \beta} \right) ^2
\left( {\sin
\theta + i (1 - \cos \theta ) \over 2} \right) ^2 \right] \right] \right\}
\label{4.2.44}
\eeq

In a Hall device, the electron effective mass and the constant homogeneous
magnetic field are given. This means that
$\beta$ is fixed and $\left( {d \lambda \over d \theta} \right) _{\rm Lab}$
is a function of two variables: $\gamma
= \omega / m$, the ratio of the energy of the incoming photon to the
electron effective mass, and $\theta$, the
scattering angle. We analyse in turn the dependence of the differential
scattering cross-length on the photon
energy for fixed $\theta$ and the angular distribution of the scattering
for fixed $\gamma$.

In order to unveil the values of $\gamma$ for which $\left( {d \lambda
\over d \theta} \right) _{\rm Lab}$ diverges,
we summmarize the properties of the confluent hypergeometric functions and
the Gamma function.
\begin{itemize}

\item A. $\ _1F_1[a,b;z]$ is:
\begin{itemize}
\item i. A convergent series for all values of $a$, $b$ and $z$ if $a \neq
- n$ and $b \neq - n'$, with $n$, $n'$
positive integers.
\item ii. A polynomial of degree $n$ in $z$ if $a = -n$ and $b\neq -n'$. $\
_1F_1[a,b;z]$ has a simple pole at
$b=-n'$ if $a=-n$, $b=-n'$ and $n > n'$. $\ _1F_1[a,b;z]$ is undefined if
$a=-n$, $b=-n'$ and $ n' \leq n$.
\end{itemize}
\item B. $\Gamma[z]$ is single-valued and analytic over the entire complex
plane, except for the points $z=-n$,
$n=0, 1,2, \cdots$ where it has simple poles.
\end{itemize}

In our formula $a=-n$ is tantamount to $\omega = - m \pm \sqrt{2 e B (n+1)
+ m^2}$ whereas $b=-n'$ requires $\omega
= - m \pm \sqrt{2 e B ( n'+ 3/2) + m^2}$. Both identities cannot happen
simultaneously for any values of $n$ and
$n'$ and there are no divergences in $\left( {d \lambda \over d \theta}
\right) _{\rm Lab}$ due to singularities of
$\ _1 F_1$. The Gamma function entering the term in $\left( {d \lambda
\over d \theta} \right) _{\rm Lab}$ due to
direct scattering has poles at the values $\omega = - m \pm \sqrt{2 e B
(n+1) + m^2}$. Thus, $\left( {d \lambda
\over d \theta} \right) _{\rm Lab}$ becomes infinite when $\omega = E_{n+1}
- E_0$, corresponding the physical
situation when the incoming photon is captured by the electron to jump to
the $(n+1)$ Landau level.  The other sign
cannot arise in this physical process ($\omega$ would be negative)
but this divergence is the signal for the
opposite phenomenon: photon emission. In the exchange term, however, the
singularities in the Gamma function appear
when $\omega = m \pm \sqrt{2 e B (n+1) +m^2}$. Again, only the plus sign for
$\omega$ is acceptable and we have a
divergent $\left( {d \lambda \over d \theta} \right) _{\rm Lab}$ when
$\omega = E_0 + E_{n+1} = E_0 - E_{-n-1}$.
The incoming photon is captured by an electron that occupies a negative energy
state in the Dirac-Landau sea and
jumps to the first Landau level. Of course, this process can be
re-interpreted as  \lq\lq pair" creation and, also,
the other sign, which is not compatible with the incoming photons, would
correspond to \lq\lq pair" annihilation of
electrons and positrons. Observe that there is a $2 m$ gap with respect to
the other \lq\lq divergences". It should
be noticed that in the interference term of $\left( {d \lambda \over d
\theta} \right) _{\rm Lab}$ the two kinds of
divergences enter together.

In short, the differential and total cross-lengths present divergences at
values of the ongoing photon energies
corresponding to the energy gaps between the lowest and the other
(positive \& negative) Dirac-Landau states. For
these energies there is no scattering but the absorption of photons and
transitions from one Dirac-Landau state to
another takes place. We encounter a phenomenon well known in the quantum theory
of radiation: resonance fluorescence. In
the scattering of light by atoms described by the Kramers-Heisenberg
formula, similar divergences appear, see
reference \cite{Sakurai}.  We are also wrongly assuming that the
life-time of the intermediate states is
infinity. These states are indeed unstable due to spontaneous emission of
photons. The energy picks an imaginary
contribution that measures the resonance width $\Gamma_r = 1 / \tau_r$, the
inverse of the life-time; replacing $E_r$ by
$E_r - i \Gamma_r/2$ in formula (\ref{4.1.38}), $c$ and $c'$ become
imaginary in such a way that the products of the
Gamma function entering in (\ref{4.2.43}) are regular and  ${d \lambda
\over d \theta}$ reaches finite maxima for
$\omega = \mp (E_0 - E_{\pm n \pm 1})$.

In practice, for other values of $\omega$ $\Gamma_r$ can be ignored. The
differential cross-length of scattering is
regular and a study of the angular distribution of $\left( {d \lambda \over
d \theta} \right) _{\rm Lab}$ is
possible. A MATHEMATICA plot of the antenna pattern encoded by formula
(\ref{4.2.44}) for  ${d \lambda \over d
\theta}$ is depicted in Figure 1 for incoming photon energies in the
ultraviolet/infrared range of the
electromagnetic spectrum. Here we are thinking of a MISFET, at $1.2$ Kelvin
degrees
of temperature and a  very low filling factor; also, $m=0.006 \ m_e$, $B/c= 6
\ {\rm Teslas}$, i.e. $L_T = 7.43
\cdot 10^{-10} \ {\rm cm}$. in the c.g.s. system.

In this range of photon energies, far from the pair creation zone, the
graphic work reveal a general pattern which
can be explained as follows:

1. $\omega < E_1- E_0$. The photon comes through the $x_2$-axis toward the
electron, which is in one state of the
$E_0$-level. The charge distribution is accelerated up and down the
$x_1$-axis in a motion of very low amplitude by
the incoming transverse electric field. The antenna pattern of the
electromagnetic field emitted by this
oscillatory shaking of the electron is similar to the same distribution in
the $B=0$ case: we find maximum
probability of photon emission in forward and backward scattering.

2. $\omega = E_1- E_0$. The ongoing photon is absorbed by the $E_0$-level
electron and a resonance in the
$E_1$-level is formed. In the excited level the electron oscillates up and
down the $x_2$-axis; recall the
$H_1(x_2-x_2^0)$ factor in the wave function. Thus, the angular
distribution of the spontaneously emitted
photons undergoes an abrupt change: there is now maximum probability of
finding the scattered photons at the
angles $\theta = 90^{\circ}$ and $270^{\circ}$ .

Before going on, notice also that the photon energy values
\[
\omega = - m  + \sqrt{(2 n_1 + 3) e B + m^2} \equiv \epsilon_{n_1}^- \ , \
n_1 = 0, 1, 2, \cdots
\]
are special. If $\omega = \epsilon_{n_1}^-$, the contribution of the direct
and interference terms to the
differential cross-length is zero; the scattering is solely due to the
exchange diagram. As a function of
$\gamma$ and $\theta$, $\left( {d \lambda \over d \theta} \right) _{\rm
Lab}$ shows saddle points at $\omega =
\epsilon_{n_1}^-$.

3. $E_1-E_0 < \omega < \epsilon_0^-$. The probability of resonance
fluorescence decreases with increasing $\omega$
in this interval. Non-resonant amplitudes become more and more important
and interfere with the resonant one. There
are two competing effects of the photon collision: first, the resonance
fluorescence induces an oscillatory motion of
the electron on the $x_2$-axis; second, the non-resonant amplitudes shake
the electron up and down the $x_1$-axis.
The more to the left on the energy interval the more preponderant is the
first movement over the second. Thus,
$90^{\circ}$ and $270^{\circ}$ are favoured, although the maxima are
flattened throughout the interval from left to
right. It is amusing to note that for these energies photons scatter out of
electrons in a Hall device just like the
quasi-particle anyonic excitations in the quantum Hall effect do between
themselves.

4. $\epsilon_0^- \leq \omega\leq E_2-E_0$. If $\omega= \epsilon_0^-$, the
angular distribution is isotropic within
one part in a million. This is due to the perfect balance of the resonant
and non-resonant amplitudes in the direct
scattering, leaving only the contribution of the exchange graph; in
this bremsstrahlung there are no preferred directions.
Beyond this point, the non-resonant amplitudes are preponderant and the
antenna pattern in the range $\epsilon_0^- <
\omega < E_2-E_0$ is as in the $B=0$ case. When $\omega =E_2-E_0$, the
next resonance is reached and a new change
in the angular distribution appear.

Below the pair creation threshold $\omega=2 m$, this behaviour is
periodically repeated. The forward-backward and
left-right symmetries, however, cease to be almost perfect for higher values
of $\gamma$ due  to stronger quantum
fluctuations. Instead, $\theta=0$ scattering in the first regime and $\theta=\pm{\pi \over 6}$  in the second
become dominant. For lighter efective mass, this behaviour is reached
before. Figure 2 shows plots of the
differential cross-length as a function of $\gamma$ for $\theta=0$. In the
second graph the effective mass has been
chosen in such a way that the threshold for pair creation occurs at values
of $\gamma$ for soft $X$-rays. The
$\gamma =2.35$ angular distribution of photon emission is due to  pair
annihilation and thus shows a maximum at
$\theta =0$. Beyond this energy, the resonances are so short-lived that the
angular distribution does not change
when they are formed. It seems that rather than two quantum mechanical
processes       of absorption/emission, a single resonant scattering takes
place when $\omega > 2m$. There are also no changes in the antenna pattern,
either in the saddle points $\omega =
\epsilon_{n_1}^{-} $ or in another type of saddle point reached when:
\begin{equation}
\omega = m+\sqrt {(2n_2+3)eB+m^2}=\epsilon_{n_2}^{+}, n_2=0,1,2,...  .
\end{equation}
In these last points there is no contribution of the exchange diagram
to   $\left( {d \lambda \over d \theta} \right) _{\rm Lab}$   and only the
direct graph contributes to a very weak light/X-ray scattering.

Numerical integration of the differential cross-length provides us with the
total cross-length of scattering. A
picture of the function $\lambda_T (\gamma)$ is shown in Figure 3. As
expected, divergences appear at the values
of $\gamma$ that coincides with the Landau energy levels. In contrast to the
ordinary planar Compton effect, no
infrared divergence due to soft photons arises in $\lambda_T$ because the
magnetic field supplies an infrared
cut-off.

\appendix

\section{Gamma Matrices  and the Electromagnetic Field in 3-dimensional Space-time}

The Dirac (Clifford) algebra in the 3-dimensional Minkowski space $M_3 =
{\bf R}^{1,2}$ is built from the
three gamma matrices $\gamma^{\mu}$ satisfying the anticommutation relations:
\eqn
\{ \gamma^{\mu}, \gamma^{\nu}\} = 2 g^{\mu \nu} \label{A1}
\een
\[
\mu = 0, 1, 2 \q , \q g^{\mu \nu} = {\rm diag} (1, -1, -1)
\]
and the hermiticity conditions  ${\gamma^{\mu}}^{\dag} = \gamma^0
\gamma^{\mu} \gamma^0$. The tensors
\eqn
1\,, \ \gamma^{\mu}\, , \ \gamma^{\mu_1} \gamma^{\mu_2}\, , \
\gamma^{\mu_1}\gamma^{\mu_2}\gamma^{\mu_3}\, ; \
\mu_1 < \mu_2 < \mu_3 \label{A2}
\een
with respect to the ${\rm SO}(2,1)$-group, the piece connected to the
identity of the Lorentz group in flatland,
form the basis of the Dirac algebra, which is thus $2^3$-dimensional. $1$ and
$\gamma^{\mu_1}\gamma^{\mu_2}\gamma^{\mu_3}= - i
\epsilon^{{\mu_1}{\mu_2}{\mu_3}} 1$ are respectively scalar and
pseudo-scalar objects. $\gamma^{\mu}$ is a three-vector but $\gamma^{\mu_1}
\gamma^{\mu_2}$ can be seen
alternatively as a anti-symmetric tensor or a pseudo-vector, which are
equivalent irreducible representations of
the ${\rm SO}(2,1)$-group. If we denote by $\epsilon^{\mu \nu \rho}$ the
completely antisymmetric tensor, equal to
+1(-1) for an even (odd) permutation of (0,1,2) and to 0 otherwise, the
$\gamma$-matrices must also satisfy
the commutation relations:
\eqn
\sigma^{\mu \nu} = {i \over 2}[ \gamma^{\mu}, \gamma^{\nu} ] =
\epsilon^{\mu \nu \rho} \gamma^{\rho}
\label{A3}
\een
The $\sigma^{\mu \nu}$-matrices are the Lie algebra generators of the $
{\rm spin}(1,2;{\rm \bf R}) \cong {\rm
SL}(2;{\rm \bf R})$-group, the universal covering of the connected piece of
the Lorentz group and the irreducible representations of the Lie ${\rm SL}(2;{\rm \bf R})$-group are the spinors. Our choice of the representation of the Dirac algebra is as follows:
\[
\gamma^0 = \sigma^3 \ , \ \gamma^1 = i \sigma^1 \ , \ \gamma^2 = i \sigma^2
\]
where the $\sigma^a$, $a = 1, 2, 3$ are the Pauli matrices. 

The canonical quantization of the electromagnetic field in $(2+1)$-dimensions is
equivalent to the four-dimensional case. We shall follow the covariant
formalism
of Gupta and Bleuler, see \cite{Mandl}. We consider the Fermi Lagrangian
density
\eqn
{\cal L} = -{1\over 2} \left( \partial_{\nu} a_{\mu}(x) \right) \left(
\partial^{\nu} a^{\mu}(x) \right) \label{D1}
\een
where now $a^{\mu}(x) \, , \, \mu = 0, 1, 2$ is the three-vector potential. The
fields equations are
\eqn
\Box a^{\mu}(x) = 0 \label{D2}
\een
which are equivalent to Maxwell's equations if the potential satisfies the
Lorentz condition $\partial_{\mu} a^{\mu} (x) = 0$. We expand the free
electromagnetic field in a complete set of plane wave states:
\beq
a^{\mu}(x) &=& a^{\mu +}(x) + a^{\mu -}(x)  \nm \\
& = & \sum_{\vec k, r} {1 \over \sqrt{2 A \omega_{\vec k}} }
\left( \epsilon_r^{\mu}(\vec k)\, b_r(\vec k)\, e^{-i k x} +
\epsilon_r^{\mu}(\vec k)\,
b_r^{\dag} (\vec k)\, e^{i k x} \right) \label{D3}
\eeq
Here, the summation is over wave vectors, allowed by the periodic boundary
conditions in $A$, with $k^0 = {1 \over c} \omega_{\vec k} = |
\vec k|$. The summation over $r = 0, 1, 2$ corresponds to the three linearly
independent polarizations states that exist for each $\vec k$. The real
polarization vectors $\epsilon_r^{\mu} (\vec k)$ satisfy the orthonormality and
completeness relations
\beq
 & & \epsilon_{r \mu} (\vec k) \epsilon_s^{\mu}(\vec k) = - \eta_r
\delta_{rs}, \q
r, s = 0, 1, 2 \label{D4} \\
& & \sum_{r} \eta_r \epsilon_r^{\mu}(\vec k) \epsilon_r^{\nu}(\vec k) = - g^{\mu
\nu} \label{D5} \\
& &  \eta_0 = -1 \ ,  \ \eta_1 = \eta_2 = 1 \nm
\eeq

The equal-time commutation relations for the fields $a^{\mu}(x)$ and their
momenta $\pi^{\mu}(x) = -{1\over c^2} {\dot a}^{\mu}(x)$ are
\beq
& & [ a^{\mu} (\vec x, t), a^{\nu} (\vec {x'}, t) ] =  [ {\dot a}^{\mu} (\vec x,
t), {\dot a}^{\nu} (\vec {x'}, t) ] = 0 \nm \\
& & [ a^{\mu} (\vec x, t), {\dot a}^{\nu} (\vec {x'}, t) ] = - i \hbar c^2
g^{\mu
\nu} \delta^{(2)} (\vec x - \vec {x'}) \label{D6}
\eeq
The operators $b_r(\vec k) $ and $b_r^{\dag}(\vec k)$ satisfy
\eqn
[ b_r(\vec k), b_s^{\dag} (\vec {k'}) ] = \eta_r \delta_{r s} \delta_{\vec k
\vec {k'}} \label{D7}
\een
and all other commutators vanish. For each value of $r$ there are transverse $(r
=1)$, longitudinal $(r=2)$ and scalar $(r=0)$ photons, but as result of the
Lorentz
condition, which in the Gupta-Bleuler theory is replaced by a restriction on the
states, only transverse photons are observed as free particles. This is
accomplished as follows: the states of the
basis of the bosonic Fock space have the form,
\[
|n_{r_1}(\vec{k_1}) n_{r_2}(\vec{k_2}) \cdots n_{r_N}(\vec{k_N}) \rangle
\propto \left[ a_{r_1}^{\dag}
(\vec{k_1})\right] ^{n_{r_1}(\vec{k_1})}  \left[ a_{r_2}^{\dag}
(\vec{k_2})\right] ^{n_{r_2}(\vec{k_2})} \cdots  \left[ a_{r_N}^{\dag}
(\vec{k_N})\right] ^{n_{r_N}(\vec{k_N})} |0 \rangle ,
\]
where $ n_{r_i}(\vec{k_i}) \in {\bf Z}^+$, $\forall i = 1, 2, \cdots, N$ and
\[
a_r(\vec k) |0 \rangle = 0 \ , \ r = 0, 1, 2
\]
defines the vacuum state. To avoid negative norm states the condition
\[
\left[ a_2(\vec k)  - a_0 (\vec k) \right] | \Psi \rangle = 0 \ , \
\forall \ \vec k \Longleftrightarrow
\langle \Psi |N_2(\vec k) |\Psi \rangle = \langle \Psi |N_0 (\vec k) |\Psi
\rangle
\]
is required on the physical photon states of the Hilbert space. Therefore,
in two dimensions, there is only one
degree of freedom for each $\vec k$ of the radiation field.

From the covariant commutation relations we derive the Feynman photon
propagator:
\eqn
\langle 0 | T \{ a^{\mu} (x) a^{\nu} (y) \} | 0 \rangle = i \hbar c
D_F^{\mu \nu}
(x - y) \label{D8}
\een
where
\eqn
D_F^{\mu \nu}(x) = {1 \over (2 \pi )^3 } \int d^3 k {- g^{\mu \nu} \over k^2 + i
\epsilon} e^{-i k x} \label{D9}
\een

Choosing the polarization vectors in a given frame of reference as
\beq
& & \epsilon_0^{\mu}(\vec k) = n^{\mu} = (1, 0, 0) \nm \\
& & \epsilon_1^{\mu}(\vec k) = (0, {\vec \epsilon}_1(\vec k) ) \ , \ {\vec
\epsilon}_1(\vec k)  \cdot \vec k = 0 \label{D10} \\
& & \epsilon_2^{\mu}(\vec k) = (0, {\vec k \over |\vec k|}) = {k^{\mu} - (k
n) n^{\mu} \over ( (kn)^2 -
k^2)^{1/2}}  \nm
\eeq
it is possible to express the momentum space propagator from (\ref{D9})
as
\beq
D_F^{\mu \nu}(k) &=& {- g^{\mu \nu} \over k^2 + i \epsilon} \nm \\
&=& D_{FT}^{\muæ\nu}(k) + D_{FC}^{\mu \nu}(k) + D_{FR}^{\mu \nu}(k)
\label{D11} \\
&=&  {1 \over k^2 + i \epsilon} \epsilon_1^{\mu} (\vec k)
\epsilon_1^{\nu}(\vec k)
+  {n^{\mu} n^{\nu} \over (kn)^2 - k^2}  + {1\over k^2 + i \epsilon} \left[
{k^{\mu} k^{\nu} -(kn) (k^{\mu} n^{\nu} + k^{\nu} n^{\mu}) \over (kn)^2 -k^2}
\right] \nm
\eeq

The first term in (\ref{D11}) can be interpreted as the exchange of transverse
photons. The remaining two terms follow from a linear combination of
longitudinal and temporal photons such that
\eqn
D_{FC}^{\mu \nu}(x) = {g^{\mu 0} g^{\nu 0} \over (2 \pi)^3 }\int {d^2 \vec
k e^{i
\vec k \cdot \vec x} \over |\vec k|^2 } \int d k^0 e^{- i k^0 x^0} =  g^{\mu 0}
g^{\nu 0} {1\over 4 \pi } \ln {1\over |\vec x|} \delta( x^0)    ;\label{D12}
\een
This term corresponds to the instantaneous Coulomb interaction between
charges in the
plane, and the contribution of the remaining term $D_{FR}^{\mu \nu}(k)$
vanishes
because the electromagnetic field only interacts with the conserved
charge-current density, \cite{Mandl}.

\b
\b
\b

$\bf{ACKNOWLEDGEMENTS}$

The authors gratefully acknowledge discussions on the physical understanding
of the scattering length with W. Garcia-Fuertes.

\vfill\eject

\begin{figure}[htbp]
\begin{center}
\epsfig{file=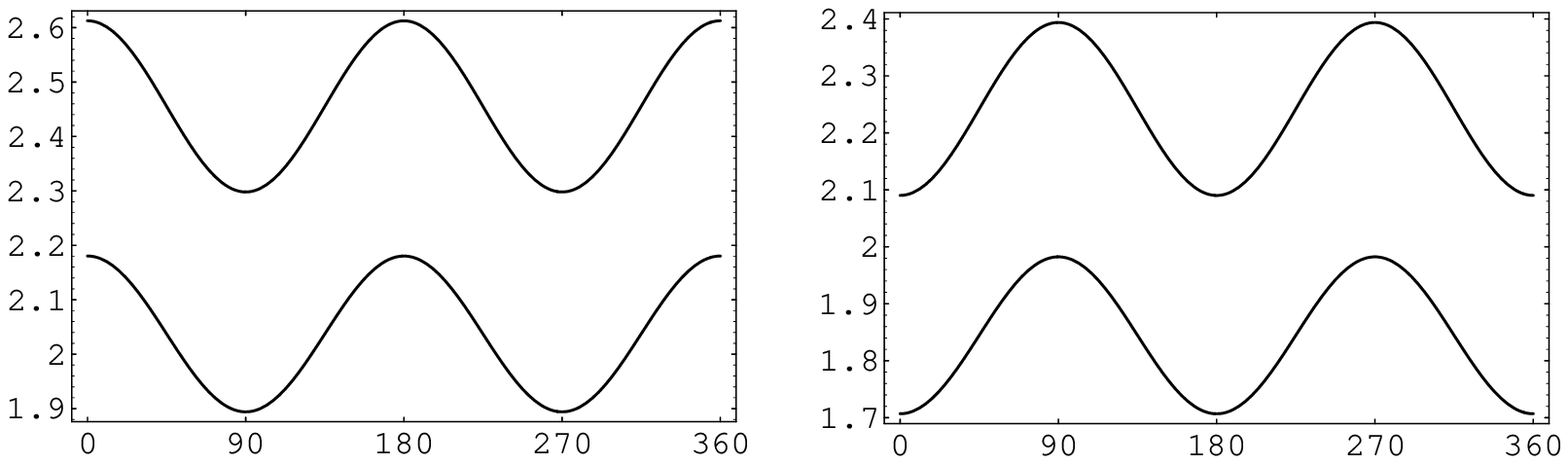,height=3cm}
\end{center}
\begin{center}
\epsfig{file=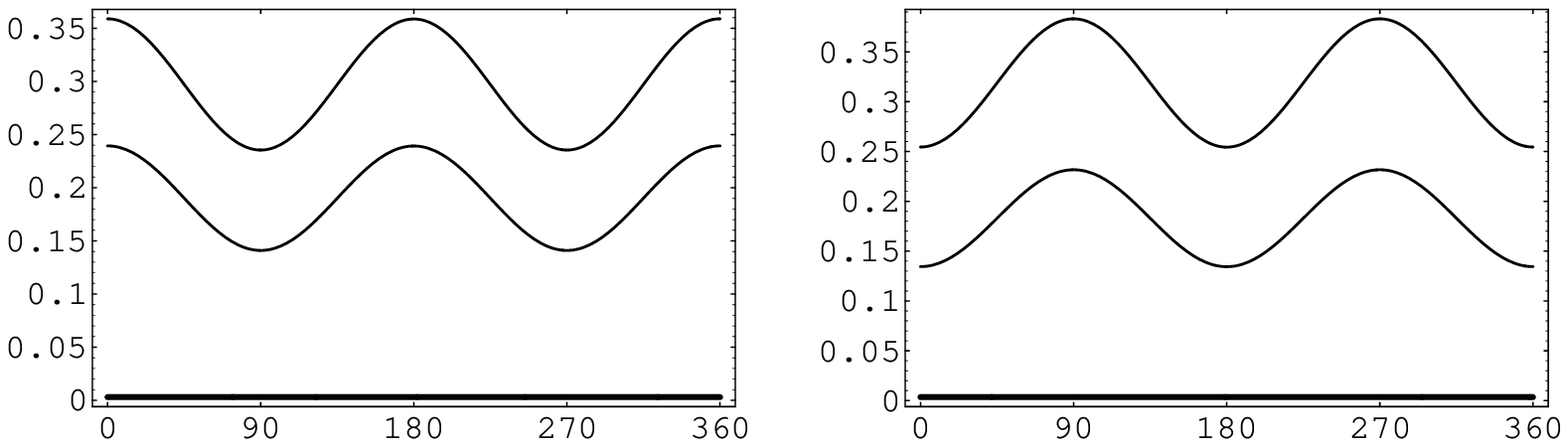,height=3cm}
\end{center}
\begin{center}
\epsfig{file=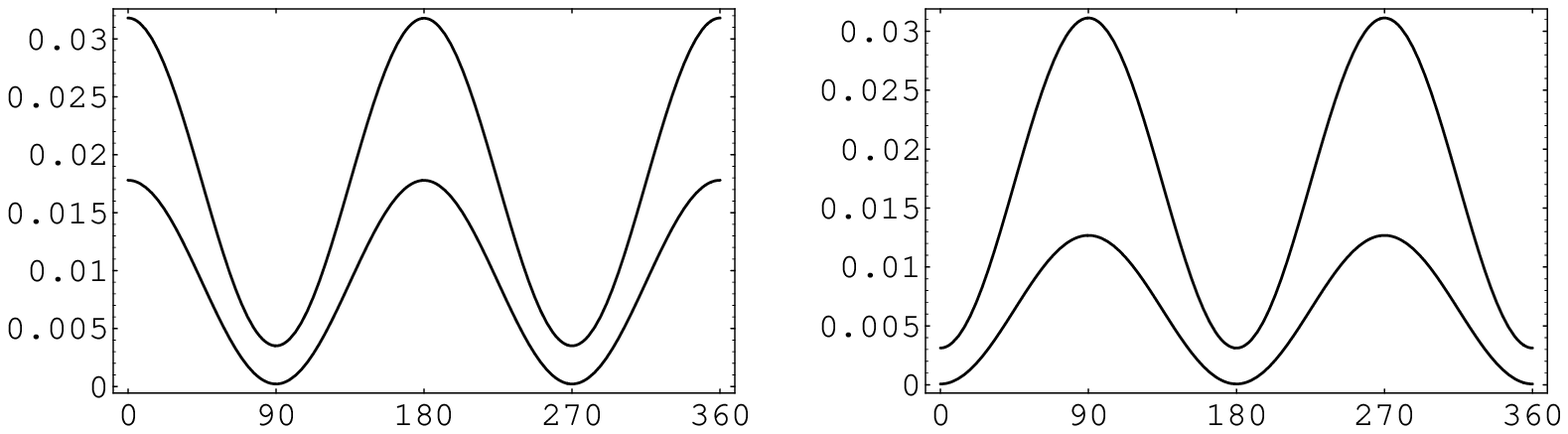,height=3cm}
\end{center}
\caption[Figure 1]{\small Angular distribution ${1\over \pi L_T} \left( {d \lambda \over d \theta} \right) $ as a function of the scattering angle $\theta$ for several values of $\gamma$: (a) In
this case, $\omega < E_1- E_0$. (b) $\omega
> E_1 -E_0$. (c) $\omega < E_2 -E_0$ and the straight line corresponds to
$\omega = \epsilon_{0}^-$ the first saddle
point. (d) $\omega > E_2 -E_0 $ and the straight line for $\omega =
\epsilon_1^-$. (e) $\omega < E_3-E_0$ and finally
(f) $\omega > E_3-E_0$. In this case, $\beta = 0.009$ and $L_T = 7.43 \cdot
10^{-10} {\rm cm}$.  We have chosen
the $[0,2\pi ]$ interval because the $ \theta \rightarrow -\theta $
symmetry is not evident from the formula.}
\end{figure}

\begin{figure}[htbp]
\begin{center}
\epsfig{file=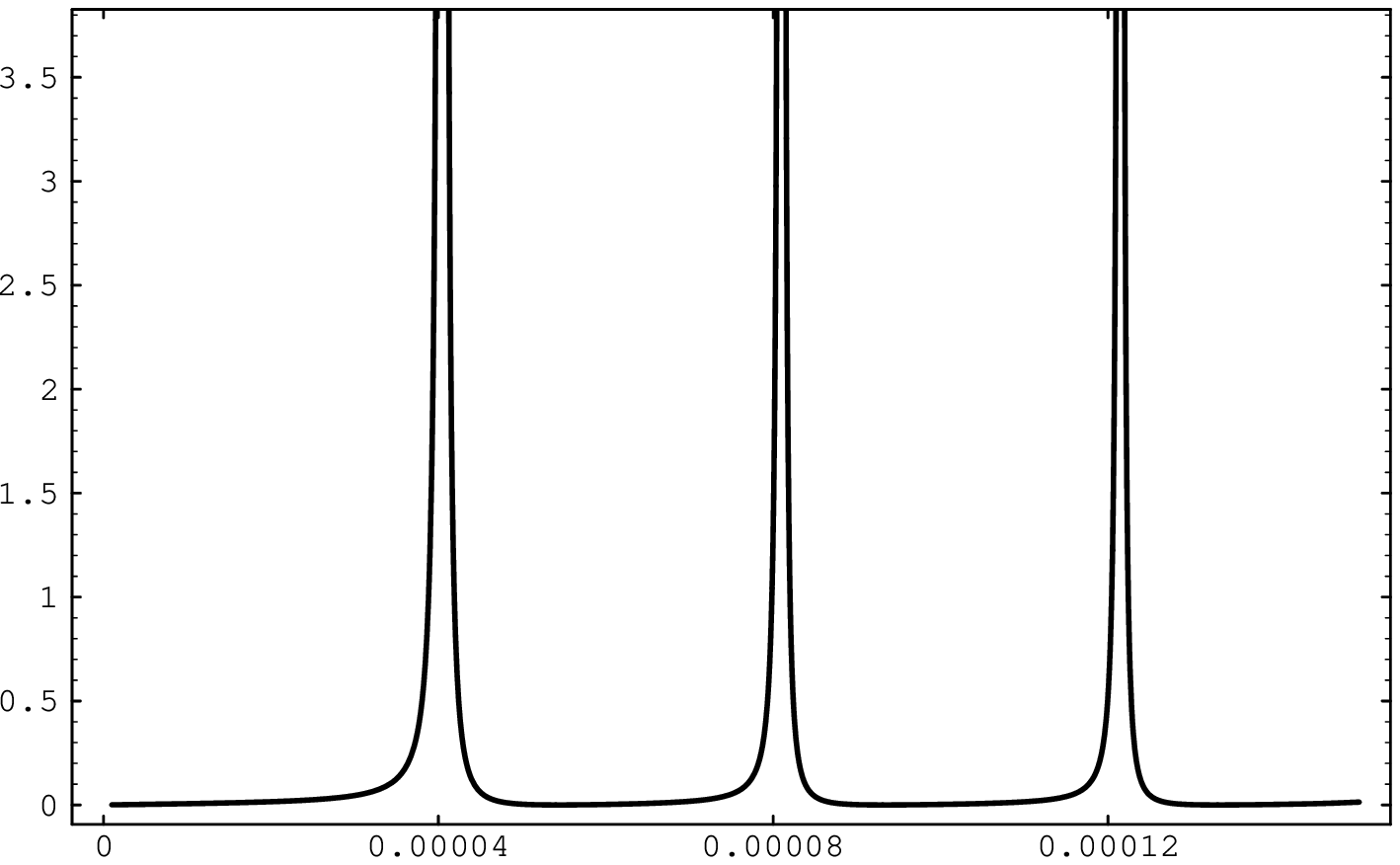,height=6cm}
\end{center}
\begin{center}
\epsfig{file=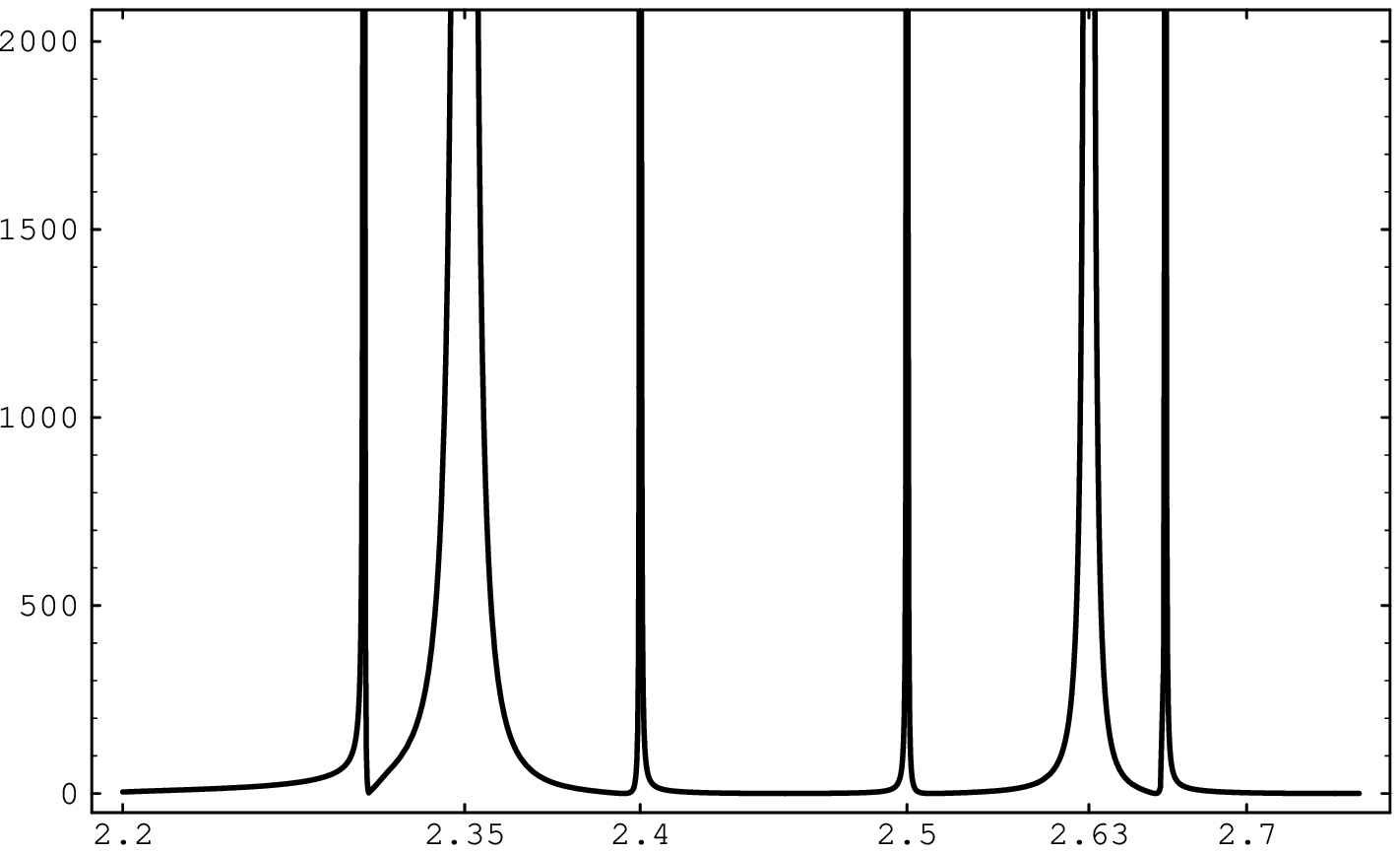,height=6cm}
\end{center}
\caption[Figure 2]{\small The differential cross-length distribution ${1\over \pi L_T} \left( {d \lambda \over d \theta} \right) $ as a
function of $\gamma$ for $\theta= 0$. (a) For $\beta = 0.009$ and $L_T = 7.43
\cdot 10^{-10} {\rm cm}$. (b) For $\beta=0.9$ and the same $L_T$.  }
\end{figure}

\begin{figure}[htbp]
\begin{center}
\epsfig{file=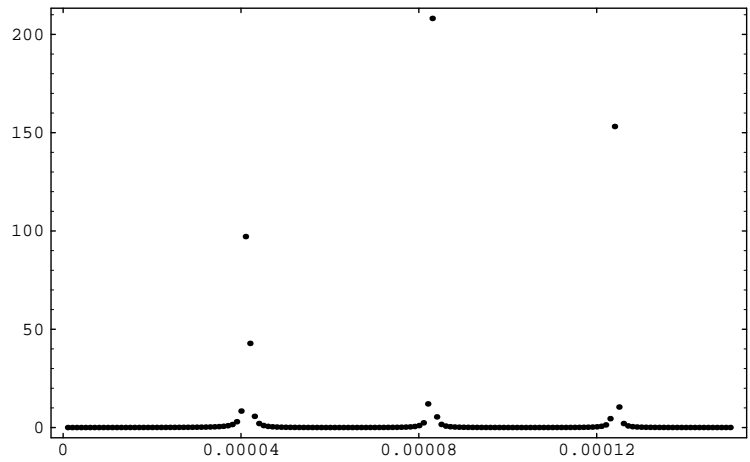,height=6cm}
\end{center}
\caption[Figure 3]{\small Total cross-length ${\lambda_T \over \pi L_T}$ as a function of $\gamma$. $\beta = 0.009$ and $L_T = 7.43
\cdot 10^{-10} {\rm cm}$.}
\end{figure}

\end{document}